\documentclass[12pt,canadian]{article}
\usepackage[latin9]{inputenc}
\usepackage[letterpaper]{geometry}
\PassOptionsToPackage{ruled, vlined, linesnumbered}{algorithm2e}
\usepackage{color}
\usepackage{babel}
\usepackage{algorithm2e}
\usepackage{amsmath}
\usepackage{amsthm}
\usepackage{graphicx}
\usepackage{grffile}
\usepackage{setspace}
\usepackage[authoryear]{natbib}
\usepackage{xargs}[2008/03/08]
\usepackage{hyperref}
\hypersetup{unicode=true,pdfusetitle,
	bookmarks=true,bookmarksnumbered=false,bookmarksopen=false,
	breaklinks=false,pdfborder={0 0 0},pdfborderstyle={},backref=false,colorlinks=true,
	linkcolor=pigment, citecolor=pigment, urlcolor=pigment}

\makeatletter

\definecolor{pigment}{rgb}{0.2, 0.2, 0.6}
\usepackage{arydshln}
\usepackage{breakcites}
\usepackage{breakurl}
\usepackage{amsfonts}
\usepackage[noblocks]{authblk}

\renewcommand\paragraph{%
        \@startsection{paragraph}{4}{0mm}%
           {-\baselineskip}%
           {.5\baselineskip}%
           {\normalfont\normalsize\bfseries}}

\author[1,*]{Samuel M. Fischer}
\author[2]{Martina Beck}
\author[3]{Leif-Matthias Herborg}
\author[1,4]{Mark A. Lewis}
\affil[1]{Department of Mathematical and Statistical Sciences, University of Alberta, Edmonton, AB.}
\affil[2]{BC Ministry of Environment and Climate Change Strategy, Conservation Science Section, Victoria, BC.}
\affil[3]{Fisheries and Oceans Canada, Institute of Ocean Sciences, Sidney, BC.}
\affil[4]{Department of Biological Sciences, University of Alberta, Edmonton, AB.}
\affil[*]{Department for Mathematical and Statistical Sciences; 632 Central Academic Building; 
University of Alberta; Edmonton, AB; T6G 2G1; E-Mail: samuel.fischer@ualberta.ca}

\date{}

\@ifundefined{showcaptionsetup}{}{%
 \PassOptionsToPackage{caption=false}{subfig}}
\usepackage{subfig}
\makeatother

\begin{document}
\global\long\def\argparentheses#1{\mathopen{\left(#1\right)}\mathclose{}}%
\global\long\def\ap#1{\argparentheses{#1}\mathclose{}}%

\global\long\def\var#1{\mathbb{V}\argparentheses{#1}}%
\global\long\def\pr#1{\mathbb{P}\argparentheses{#1}}%
\global\long\def\ev#1{\mathbb{E}\argparentheses{#1}}%

\global\long\def\smft#1#2{\stackrel[#1]{#2}{\sum}}%
\global\long\def\smo#1{\underset{#1}{\sum}}%
\global\long\def\prft#1#2{\stackrel[#1]{#2}{\prod}}%
\global\long\def\pro#1{\underset{#1}{\prod}}%
\global\long\def\uno#1{\underset{#1}{\bigcup}}%

\global\long\def\order#1{\mathcal{O}\argparentheses{#1}}%
\global\long\def\R{\mathbb{R}}%
\global\long\def\Q{\mathbb{Q}}%
\global\long\def\N{\mathbb{N}}%
\global\long\def\Z{\mathbb{Z}}%
\global\long\def\F{\mathcal{F}}%

\global\long\def\mathtext#1{\mathrm{#1}}%
\global\long\def\mt#1{\mathtext{#1}}%

\global\long\def\maxo#1{\underset{#1}{\max\,}}%
\global\long\def\argmaxo#1{\underset{#1}{\mathtext{argmax}\,}}%
\global\long\def\minargmaxo#1{\underset{#1}{\mathtext{minargmax}\,}}%
\global\long\def\argsupo#1{\underset{#1}{\mathtext{argsup}\,}}%
\global\long\def\supo#1{\underset{#1}{\sup\,}}%
\global\long\def\info#1{\underset{#1}{\inf\,}}%
\global\long\def\mino#1{\underset{#1}{\min\,}}%
\global\long\def\argmino#1{\underset{#1}{\mathtext{argmin}\,}}%
\global\long\def\limo#1#2{\underset{#1\rightarrow#2}{\lim}}%
\global\long\def\supo#1{\underset{#1}{\sup}}%
\global\long\def\info#1{\underset{#1}{\inf}}%

\global\long\def\b#1{\boldsymbol{#1}}%
\global\long\def\ol#1{\overline{#1}}%
\global\long\def\ul#1{\underline{#1}}%

\newcommandx\der[3][usedefault, addprefix=\global, 1=, 2=, 3=]{\frac{d^{#2}#3}{d#1^{#2}}}%
\newcommandx\pder[3][usedefault, addprefix=\global, 1=, 2=]{\frac{\partial^{#2}#3}{\partial#1^{#2}}}%
\global\long\def\intft#1#2#3#4{\int\limits _{#1}^{#2}#3d#4}%
\global\long\def\into#1#2#3{\underset{#1}{\int}#2d#3}%

\global\long\def\th{\theta}%
\global\long\def\la#1{~#1~}%
\global\long\def\laq{\la =}%
\global\long\def\normal#1#2{\mathcal{N}\argparentheses{#1,\,#2}}%
\global\long\def\uniform#1#2{\mathcal{U}\argparentheses{#1,\,#2}}%
\global\long\def\I#1#2{\mbox{I}_{#1}\argparentheses{#2}}%
\global\long\def\chisq#1{\chi_{#1}^{2}}%
\global\long\def\dar{\,\Longrightarrow\,}%
\global\long\def\dal{\,\Longleftarrow\,}%
\global\long\def\dad{\,\Longleftrightarrow\,}%
\global\long\def\norm#1{\left\Vert #1\right\Vert }%
\global\long\def\code#1{\mathtt{#1}}%
\global\long\def\descr#1#2{\underset{#2}{\underbrace{#1}}}%
\global\long\def\NB{\mathcal{NB}}%
\global\long\def\BNB{\mathcal{BNB}}%
\global\long\def\e#1{\text{e}_{#1}}%
\global\long\def\P{\mathbb{P}}%
\global\long\def\pb#1{\Bigg(#1\Bigg)}%
\global\long\def\cpb#1{\Big[#1\Big]}%

\global\long\def\vv#1{\boldsymbol{#1}}%

\global\long\def\True{\mathtt{True}}%

\global\long\def\False{\mathtt{False}}%

\newgeometry{top=0cm, left=1.5cm, right=1.5cm, bottom=2cm}
\title{Managing Aquatic Invasions: Optimal Locations and Operating Times
for Watercraft Inspection Stations}
\maketitle
\begin{abstract}
Aquatic invasive species (AIS) cause significant ecological and economic
damages around the world. A major spread mechanism for AIS is traffic
of boaters transporting their watercraft from invaded to uninvaded
waterbodies. To inhibit the spread of AIS, several Canadian provinces
and American states set up watercraft inspection stations at roadsides,
where potentially infested boats are screened for AIS and, if necessary,
decontaminated. However, since budgets for AIS control are limited,
watercraft inspection stations can only be operated at specific locations
and daytimes. Though theoretical studies provide managers with general
guidelines for AIS management, more specific results are needed to
determine when and where watercraft inspections would be most effective.
This is the subject of this paper. We show how linear integer programming
techniques can be used to optimize watercraft inspection policies
under budget constraints. We introduce our approach as a general framework
and apply it to the prevention of the spread of zebra and quagga mussels
(\textit{Dreissena spp}.) to the Canadian province British Columbia.
We consider a variety of scenarios and show how variations in budget
constraints, propagule sources, and model uncertainty affect the optimal
policy. Based on these results, we identify simple, generally applicable
principles for optimal AIS management.
\end{abstract}
\begin{description}
\item [{Keywords:}] aquatic invasive species; linear integer programming;
optimal management; spatially explicit; zebra mussel
\end{description}
\restoregeometry

\section{Introduction}

Human traffic and trade are major vectors for invasive species \citep{lockwood_invasion_2013}.
 Due to the significant ecological and economic damages invasive
species cause \citep{pimentel_update_2005}, government regulations
restrict the import of certain goods and require treatment of potentially
infested freight and carriers \citep{shine_assessment_2010,johnson_invasive_2017,turbelin_mapping_2017}.
While such regulations may be enforced comparatively easily at ports,
air ports, and border crossings, control of inland traffic is more
difficult, as a vast number of routes need to be monitored. This applies
for example to the spread of zebra and quagga mussels (\emph{Dreissena
spp.}) and other aquatic invasive species (AIS), which often spread
with watercraft and equipment transported from invaded to uninvaded
waterbodies \citep{johnson_overland_2001}. Zebra and quagga mussels
are invasive in North America and have negative effects on native
species and ecosystems, water quality, tourism, and infrastructure
\citep{rosaen_costs_2012,karatayev_zebra_2015}.

To counteract the spread of these AIS, watercraft inspection stations
are set up on roads, where transported watercraft are inspected for
AIS and decontaminated if at risk for carrying AIS \citep{mangin_100th_2011,alberta_environment_and_parks_fish_and_wildlife_policy_alberta_2015,inter-ministry_invasive_species_working_group_zebra_2015}.
However, since budgets for inspections are limited, not all pathways
can be monitored around the clock, and managers need to prioritize
certain locations and daytimes. Though several theoretical studies
provide managers with helpful guidelines for their work \citep{leung_ounce_2002,potapov_allee_2008,potapov_optimal_2008,vander_zanden_management_2008,finnoff_control_2010,hyytiainen_optimization_2013},
more specific results are needed in practice to determine the locations
and times where and when control is most effective.  To date it
has been difficult to tackle these questions rigorously, as comprehensive
models for road traffic of potential vectors were missing. Therefore,
AIS managers have relied on past watercraft inspection data, shared
experience between jurisdictions, and iterative improvements of control
policies. Recent modelling advances \citep{fischer_hybrid_2019},
however, now permit the application of quantitative methods to optimize
control measures in road networks and to evaluate their effectiveness.
This will be the subject of this paper.

Our goal will be to minimize the number of boaters reaching uninvaded
waterbodies without being inspected for AIS. Thereby, we will assume
that a fixed budget is available for AIS control. This problem setup
differs from scenarios considered in other studies on optimal control
of invasive species \citep{hastings_simple_2006,potapov_allee_2008,potapov_optimal_2008,finnoff_control_2010,epanchin-niell_optimal_2012},
where budget allocation over time is optimized along with the control
actions. However, to optimize the budget, invasions need to be assigned
``cost labels''. This is an often difficult and politically sensitive
task. Furthermore, the budget available for AIS control may be subject
to political and social influences and determined on a different decision
hierarchy than the management actions. Therefore, AIS managers may
seek to spend a fixed yearly budget optimally rather than to determine
the theoretically best control budget. The presence of fixed budget
constraints also reduces the need to consider the invasion as a dynamic
process.

Identifying the locations where a maximal number of boaters could
be screened for AIS is similar to the problem of finding optimal locations
for road-side infrastructure \citep{trullols_planning_2010}. A well-known
technique to solve such problems is linear integer programming \citep{conforti_integer_2014}.
The idea is to model the optimization problem with functions linear
in the decision variables. Though solving linear integer programs
is a computationally difficult task in general, good approximate solutions
can often be determined, and a variety of software tools are available
to compute solutions. Therefore, linear integer programming has also
been used in the context of invasive species management \citep{epanchin-niell_optimal_2012,kibis_optimizing_2017}.

A crucial step in linear integer optimization is to find a problem
formulation that facilitates good approximations \citep{ageev_pipage_2004}.
 In this paper, we provide such a formulation to optimize locations
and operating times of watercraft inspection stations. This problem
differs from comparable resource allocation problems \citep{surkov_model_2008,trullols_planning_2010},
as we need to account for the temporal variations in traffic. These
variations are key when we consider the trade-off between operating
few inspection stations intensely, e.g. around the clock, and distributing
resources over many locations operated at peak traffic times only.

We demonstrate the potential of our approach by applying it to optimize
watercraft inspection policies for the Canadian province British Columbia
(BC). We show how uncertainty, different cost constraints, and additional
propagule sources impact the optimal policy. Thereby, we identify
control principles applicable beyond the considered scenario.

This paper is structured as follows: we start by introducing model
components required to optimize watercraft inspection station operation.
Then, we show how the considered optimization task can be formulated
as linear integer problem. Thereby, we focus first solely on inspection
station placement before we introduce the full problem, in which also
operating times of inspection stations are optimized. After this general
description of our approach, we apply the method to AIS management
in BC and present results under different scenarios. Lastly, we discuss
our results and the limitations of our approach and draw general conclusions
on AIS management.

\section{Method}

\subsection{Model\label{subsec:Model}}

Our goal is to identify how limited resources can be allocated most
effectively to minimize the number of boaters arriving at uninvaded
waterbodies without being inspected for AIS. We assume that two aspects
of the control strategy can be changed: the locations and operating
times of watercraft inspection stations. As traffic typically follows
cyclic patterns, we consider one such cycle as the time horizon for
the control optimization.

\begin{figure}
\begin{centering}
\includegraphics{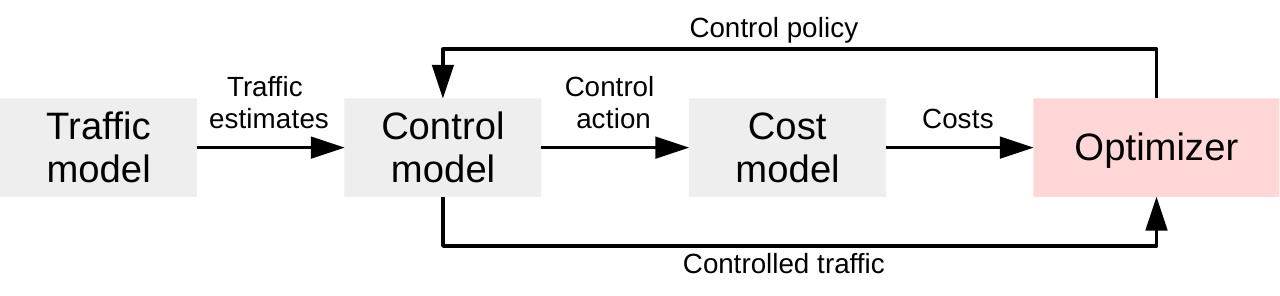}
\par\end{centering}
\caption[Components of our approach]{Components of our approach. The control model determines how the
traffic estimated by the traffic model changes under a given control
policy. The cost model yields the costs for control actions. The optimizer
maximizes the controlled traffic subject to a cost constraint.\label{fig:Flow-Chart}}
\end{figure}

To find an optimal inspection policy, we need three models (see Figure
\ref{fig:Flow-Chart}): (1) a model for boater traffic, (2) a model
for control, and (3) a model for control costs. The traffic model
gives us estimates of when, where, and along which routes boaters
travel. The control model shows us when and where inspections could
be conducted and how effective they are. Lastly, the cost model measures
the costs for inspections. The information from the three models serve
as input for a control optimizer that determines a good -- or, if
possible, the best -- watercraft inspection strategy. Below, we describe
each of the models in greater detail before we introduce suitable
optimization routines in the next section.

\subsubsection{Traffic model}

The traffic model provides estimates of when, where, and along which
routes boaters drive. Knowledge about routes is key to understanding
whether boaters passing one control location have already been inspected
at another location. For each considered route, the traffic model
provides us with a traffic estimate. In this study, we use a hybrid
gravity and route choice model \citep{fischer_hybrid_2019} to estimate
the traffic. The model includes components accounting for boaters'
travel incentive, their route choice, the timing of traffic, and boaters'
compliance with inspections.

In practice it is rarely feasible to consider all routes that boaters
could possibly take, and we need to focus on some set of ``reasonable''
routes \citep{bovy_modelling_2009,fischer_locally_2019}. As a consequence,
there may be some agents travelling along unexpected routes. When
boaters travelling along such routes arrive at inspection locations,
we do not know whether their watercraft have been inspected earlier.
This makes it difficult to optimize inspection strategies. Nonetheless,
we may want to account for these boaters by introducing a ``noise''
term to our model. To that end, we assume that a fraction of the travelling
boaters could be observed at any inspection location with a small
probability \citep[see][]{fischer_hybrid_2019}.

As road traffic is rarely uniform over time, we furthermore need a
submodel predicting how traffic varies with time. While it may be
comparatively easy to determine the temporal distribution of traffic
at a specific location, it can be difficult to identify the temporal
relationship between traffic at two locations on the same route. For
example, agents passing one location in the morning may not be able
to reach another location before the afternoon. Modelling such relationships
is particularly difficult for locations far from each other, as boaters
may have different travel speeds. We therefore apply a simplification
and assume all boaters travelling along a route have the same speed.

\subsubsection{Control model\label{subsec:Control-model}}

We assume that there is a specific set of locations where watercraft
inspections could be conducted. For example, these locations could
be pullouts large enough to provide a safe environment for inspections.
We suppose that compliant boaters stop for an inspection whenever
they pass an operated inspection station. Conversely, uncompliant
boaters are assumed to bypass any inspection station on their route.
Consequently, we seek to maximize the number of boaters that pass
at least one operated watercraft inspection.

As with the inspection locations, we assume that there are specific
time intervals when inspection can be conducted. The admissible time
intervals may be determined by safety concerns or practical considerations
and can be location dependent. As staff cannot move between distant
inspection locations easily, and the working hours of inspection staff
are subject to legal and practical constraints, we may furthermore
assume that every inspection station can be operated in shifts of
given lengths only.

\subsubsection{Cost model}

Inspection costs may be split in two classes: infrastructure costs
that apply once for each chosen inspection location, and operational
costs, which depend on when and for how long an inspection station
is operated. The operational costs may also account for ongoing equipment
maintenance costs and training of staff. The control costs may be
location and time dependent. For example, it may be expensive to conduct
inspections at remote locations if staff must travel long distances
to their work place. Furthermore, some locations will require significantly
more infrastructure costs (e.g. lighting and washrooms) in order to
operate overnight shifts. In addition, wages are often higher in overnight
shifts.

\subsection{Optimizing control locations\label{subsec:Optimizing-control-locations}}

With the submodels from the previous section at hand, we can proceed
optimizing the inspection strategy. Optimizing both locations and
operating times of watercraft inspection stations at the same time
is conceptually and computationally challenging. To ease the introduction
of our approach, we first consider a scenario in which inspection
stations are operational around the clock. In this case, we can ignore
the temporal variations of traffic and focus on choosing optimal control
locations \citep[cf. ][]{trullols_planning_2010}.

In this section, we show how the corresponding optimization problem
can be formulated as a linear integer problem. To that end, we let
$L$ be the set of all admissible inspection locations and introduce
for each location $l\in L$ a binary variable $x_{l}$ that assumes
the value $1$ if and only if an inspection station is set up at $l$.
Let $R$ be the set of potential routes that boaters may choose, $n_{r}$
the expected number of complying boaters travelling along route $r\in R$,
and $L_{r}\subseteq L$ the set of locations where the boaters travelling
on route $r$ could be inspected.

As noted earlier, one inspection station suffices to control all complying
boaters driving along a route $r$. Consequently, boaters travelling
on route $r$ will be controlled if and only if 
\begin{eqnarray}
\smo{l\in L_{r}}x_{l} & \geq & 1.\label{eq:flow-constraint}
\end{eqnarray}
Otherwise, the left hand side of equation $\eqref{eq:flow-constraint}$
will be $0$. Therefore, we can express the total number of inspected
boaters by 
\begin{eqnarray}
F_{\mt{loc}}\ap{\vv x} & := & \smo{r\in R}\min\left\{ 1,\smo{l\in L_{r}}x_{l}\right\} n_{r}.\label{eq:taget-fun}
\end{eqnarray}

To formulate the cost constraint, let $c_{l}$ be the cost for operating
control location $l\in L$ and $B$ the available budget. As we assume
that all inspection stations are operated for the same time, we do
not need to distinguish between infrastructure and operation costs.
Hence, we can write the cost constraint as 
\begin{eqnarray}
\smo{l\in L}c_{l}x_{l} & \leq & B.\label{eq:cost-constr}
\end{eqnarray}

The optimal placement policy can be identified by maximizing $F_{\mt{loc}}\ap{\vv x}$
over all $\vv x\in\left\{ 0,1\right\} ^{\left|L\right|}$ subject
to constraint (\ref{eq:cost-constr}). Though $F_{\mt{loc}}$ contains
a ``minimum'' function, $F_{\mt{loc}}$ can be easily transformed
to a linear function by introducing further variables and linear inequality
constraints (see e.g. \citealp{cornuejols_approximation_1999}). Since
the left hand side of the cost constraint (\ref{eq:cost-constr})
is linear in $\vv x$ as well, and $\vv x$ is constrained to be a
vector of integers, the considered optimization problem is a linear
integer problem. This can be solved with a suitable general linear
integer programming solver or a specifically tailored rounding algorithm
\citep{ageev_pipage_2004}. We discuss possible optimization routines
in section \ref{subsec:Solving-the-optimization-problems}.

\subsection{Optimizing control locations and timing\label{subsec:Optimizing-control-locations-timing}}

After focusing on inspection station placement, we now extend our
approach to permit free choice of inspection station operating times.
In this extended scenario, we need to balance the trade-off between
operating few highly frequented inspection stations around the clock
and distributing efforts over many locations operated at peak traffic
times only. This trade-off makes combined optimization of location
choice and timing more challenging than separate optimization of location
choice and timing \citep[cf.][]{epanchin-niell_optimal_2012}.

While location choice is a discrete optimization problem -- each
potential inspection location is either chosen or not -- optimization
of operating times is a continuous problem, since inspections could
be started at any time. To exploit the toolset of discrete optimization
anyway, we simplify our problem by discretizing time. That is, we
split the boater traffic corresponding to boaters' departure times
and consider only discrete sets of admissible inspection shifts.

Let $T$ be a set of disjunct time intervals covering the complete
time span of interest. We write $n_{rt}$ for the expected number
of boaters who travel on route $r\in R$, start their journey in time
interval $t\in T$, and are willing to comply with inspections. Let
furthermore $S_{l}$ be the set of admissible inspection shifts for
location $l\in L$. Each shift corresponds to a time interval in which
the inspection station is operated. Since the shift lengths are given,
the set $S_{l}$ can be fully characterized by the shifts' start times.

As we assume that all boaters travelling along a route have the same
speed, we can determine the set $S_{lrt}\subseteq S_{l}$ of control
shifts during which boaters who started their journey in time interval
$t\in T$ arrive at location $l\in L$ via route $r\in R$. Under
reasonable error allowance, it is usually possible to construct the
sets $S_{lrt}$ in a way that each shift covers the departure time
intervals either completely or not at all, respectively. This setup
prevents issues arising if some intervals overlap only partially.

To formulate our optimization problem as linear integer problem, we
describe the control policy again with binary variables $x_{ls}\in\left\{ 0,1\right\} $.
Here, $x_{ls}$ is $1$ if and only if an inspection station at location
$l\in L$ is operated in shift $s\in S_{lrs}$. Agents travelling
on route $r\in R$ who departed in time interval $t\in T$ are controlled
if and only if

\begin{eqnarray}
\smo{l\in L_{r}}\smo{s\in S_{lrt}}x_{ls} & \geq & 1.\label{eq:flow-contraint-1}
\end{eqnarray}
Consequently, the total controlled agent flow is given by
\begin{eqnarray}
F_{\mt{full}}\ap{\vv x} & := & \smo{r\in R}\smo{t\in T}\min\left\{ 1,\smo{l\in L_{r}}\smo{s\in S_{lrt}}x_{ls}\right\} n_{rt}.\label{eq:taget-fun-operation}
\end{eqnarray}

To derive the cost constraint, recall that we distinguish between
infrastructure costs $c_{l}^{\mt{loc}}$ for using location $l$ and
operating costs $c_{ls}^{\mt{shift}}$ payable per control shift $s$
conducted at $l$. Consequently, the total costs for control at $l$
are given by
\begin{equation}
\smo{s\in S_{l}}c_{ls}^{\mt{shift}}x_{ls}+c_{l}^{\mt{loc}}\maxo{r\in R,\,t\in T}\left(\smo{s\in S_{lrt}}x_{ls}\right),\label{eq:costs-location}
\end{equation}
and the cost constraint reads
\begin{eqnarray}
\smo{l\in L}\left(\smo{s\in S_{l}}c_{ls}^{\mt{shift}}x_{ls}+c_{l}^{\mt{loc}}\maxo{r\in R,\,t\in T}\left(\smo{s\in S_{lrt}}x_{ls}\right)\right) & \leq & B.\label{eq:cost-constr-operation}
\end{eqnarray}
As in the previous section, $B$ denotes the available budget. Optimizing
$F_{\mt{full}}$ subject to (\ref{eq:cost-constr-operation}) is a
linear integer problem, since the ``minimum'' term in (\ref{eq:taget-fun-operation})
and the ``maximum'' terms in (\ref{eq:cost-constr-operation}) can
be replaced by introducing correspondingly constrained variables.

\subsection{Noise\label{subsec:Noise}}

Even if the traffic model accounts for most routes boaters use, some
boaters may travel along unexpected routes. It is difficult to optimize
inspection station operation with regards to these boaters, as we
do not know which inspection stations cover the same routes. Nonetheless,
it can be desirable to account for noise, since the level of uncertainty
may affect the optimal inspection policy.

In the absence of a mechanistic model for traffic noise, we may assume
that boaters who are travelling on unexpected routes are passing any
inspection location with a small probability $\eta_{o}$, whereby
they choose the passing time randomly. Under this assumption, the
expected number of inspected boaters travelling along unknown routes
is given by 
\begin{eqnarray}
F_{\mt{noise}} & = & \left(1-\pro{l\in L}\left(1-\eta_{o}\smo{s\in S_{l}}x_{ls}\tau_{sl}\right)\right)n_{\mt{noise}}.\label{eq:F_noise}
\end{eqnarray}
Here, $n_{\mt{noise}}$ denotes the expected number of boaters travelling
on unknown routes.

As $F_{\mt{noise}}$ is not a convex function, adding this noise term
to the objective function would make optimization difficult. However,
as $\eta_{0}$ is typically small, equation (\ref{eq:F_noise}) is
well approximated by 
\begin{eqnarray}
\hat{F}_{\mt{noise}} & = & \eta_{o}n_{\mt{noise}}\smo{l\in L}\smo{s\in S_{l}}x_{ls}\tau_{sl},\label{eq:F_noise_approx}
\end{eqnarray}
which is linear and can thus be easily added to the linear integer
problem. This approximation is most precise if $x_{ls}=0$ for most
$l$ and $s$. If the budget is high enough to operate many inspection
stations for long times, the noise may be overestimated. However,
since $n_{\mt{noise}}$ is typically small compared to the total boater
traffic, inaccuracies in the noise model are unlikely to alter the
overall optimization results significantly.

\subsection{Solving the optimization problems\label{subsec:Solving-the-optimization-problems}}

Having derived the problem formulation in the previous sections, we
now proceed by discussing suitable solution methods. The inspection
station placement problem described in section \ref{subsec:Optimizing-control-locations}
is equivalent to the budgeted maximum coverage problem \citep{khuller_budgeted_1999},
also called maximum coverage problem with knapsack constraint \citep{ageev_pipage_2004}.
This problem is well studied in computing science, and it has been
shown that finding a solution better than factor $\left(1-e^{-1}\right)$
of the optimum is an NP-hard, often infeasibly difficult, problem
\citep{feige_threshold_1998}. This result applies also to the extended
problem introduced in section \ref{subsec:Optimizing-control-locations-timing},
as it is more general than the placement problem. Though these theoretical
results show that scenarios exist in which the problems considered
in this paper cannot be solved exactly within reasonable time, good
approximate or even optimal solutions can often be obtained in practical
applications.

When seeking a good solution, we can exploit that the linear integer
formulation of our problem helps us to obtain upper and lower bounds
to solutions efficiently. Consider a slightly changed optimization
problem in which the management variables $\vv x$ are not constrained
to be integers but drawn from the continuous domain $\left[0,1\right]^{N}$.
Here, $N$ is the dimension of the problem. In this case, the problems
can be solved with linear programming techniques within seconds even
if $N$ is large. Clearly, the integer domain $\left\{ 0,1\right\} ^{N}$
is a subset of the continuous domain $\left[0,1\right]^{N}$. Therefore,
the solution to the problem with relaxed integer constraint is an
upper bound to the desired integer solution.

Often it is possible to obtain good integer solutions by rounding
the solution to the continuous problem. \citet{ageev_pipage_2004}
present an algorithm that always achieves the approximation bound
$\left(1-e^{-1}\right)$ for the inspection station placement problem,
in which operating times are fixed. Nonetheless, general solvers with
possibly poorer worst-case performance may yield better solutions
in ``benign'' cases. A number of generally applicable methods exist
\citep{conforti_integer_2014}. In this study, we use branch and bound
methods, in which the distance between upper and lower bounds on the
optimal objective are found by solving continuously relaxed subproblems
with some constrained variables.

A challenge that general solvers face is to find a good initial feasible
solution that they can improve on. For the pure inspection station
placement problem, we could apply the rounding algorithm by \citet{ageev_pipage_2004},
which would also guarantee us the best approximation bound. However,
for the joint optimization of both placement and operating times of
watercraft inspection station, we are not aware of any algorithm with
such a guarantee. We therefore propose a ``greedy'' rounding algorithm
to obtain good initial solutions. The idea is to solve the relaxed
linear programming problem and to determine the largest non-integer
decision variable that can be rounded up without violating the cost
constraint. We applied this procedure with some improvements described
in Supplementary Appendix A. In applications, we consistently obtained
solutions better than $80\%$ of the optimum with this approach.

\section{Application}

To show the potential of our approach, we applied it to optimize watercraft
inspections in the Canadian province British Columbia (BC).  Below
we provide an overview of the scenario-specific submodels we used.
Furthermore, we briefly describe our implementation of the presented
approach.

\subsection{Scenario-specific submodels}

\subsubsection{Traffic model}

To model boater traffic, we used the hierarchical gravity and route
choice model for boater traffic presented in \citet{fischer_hybrid_2019}.
The model was fitted to data collected at British Columbian watercraft
inspection stations in the years $2015$ and $2016$. At the time
this study was conducted, dreissenid mussels were not known to be
established anywhere in BC. As sources of potentially infested boaters,
we therefore considered the Canadian provinces and American states
that (1) were known to be invaded by dreissenid mussels  or (2) had
connected waterway to an infested jurisdiction and no coordinated
mussel detection program in place at the time the data were collected.
As sinks we identified $5981$ potentially boater accessible lakes
in BC.

To estimate the boater traffic between an origin and destination,
the model considered characteristics of the donor jurisdiction, the
recipient lake, and the distance between the two. Major sources of
high-risk boaters were characterized by high population counts. Furthermore,
Canadian provinces were found to have higher boater traffic to BC
than American states. Attractiveness of destination lakes increased
with their surface area, the population counts of surrounding towns
and cities, and the availability of close-by touristic facilities,
such as campgrounds. Lastly, the boater flow was estimated to decay
in cubic order of the distance between an origin and a destination.
For a detailed description of the model along with precise parameter
estimates, refer to \citet{fischer_hybrid_2019}.

To identify potential boater pathways, we computed locally optimal
routes \citep{fischer_locally_2019} between the considered origins
and destinations. These routes arise if routing decisions on local
scales are rational and based on simple criteria (here: minimizing
travel time) whereas unknown factors may affect routing decisions
on larger scales. Consequently, the model accounts for routes arising
from a multitude of mechanisms. The attractiveness of the routes was
computed based on their length measured in travel time. Again, a more
in-depth description of the model and the fitted parameter values
can be found in \citet{fischer_hybrid_2019}.

The fraction of boaters travelling on routes not covered by our traffic
model was estimated as $4.9\%$. However, this number is not estimable
from survey data obtained at watercraft inspection stations, because
it is negatively correlated with the parameter $\eta_{o}$ (section
\ref{subsec:Noise}), denoting the probability to observe a boater
travelling on an unknown route at an arbitrary inspection location.
Therefore, \citet{fischer_hybrid_2019} introduced an additional model
assumption bounding the noise term below $5\%$. Note that due to
the dependency of $\eta_{o}$ on the noise level, the estimability
issue has little effect on the noise level observed at watercraft
inspection stations and thus on inspection policy. Based on a noise
level of $4.9\%$, $\eta_{o}$ was estimated as $0.06$ \citep{fischer_hybrid_2019}.

The temporal distribution of traffic was modelled with a von Mises
distribution. This is a unimodal circular distribution often used
in models \citep{lee_circular_2010}. The temporal pattern was assumed
to have a period of one day. The traffic high was estimated to be
at $2\,\text{PM}$, whereby the estimated peak traffic was $15$ times
higher than the estimated traffic volume at night. As traffic data
were available for specific inspection locations only, we assume that
the temporal traffic distribution is uniform over all locations.

Assuming an equal temporal traffic distribution for all potential
inspection locations makes it difficult to account for the time boaters
need to travel between two sites. This, is a model limitation but
not of major concern in the considered scenario of boater traffic
to BC. First, note that we seek locations that are \emph{not} on the
same pathway. If boaters do not pass multiple operated inspection
locations, we are safe to neglect the travel time between sites. Furthermore,
we can exploit that the considered boater origins are located outside
of the province and boaters drive, with minor exceptions, along highways
in one particular direction. Consequently, the temporal traffic distribution
of close-by locations on such a highway would be equal up to a shifting
term, and the optimized inspection times could be adjusted accordingly.

\subsubsection{Control model}

As described in section \ref{subsec:Control-model}, we assume that
every complying boater passing an operated inspection location is
inspected for invasive mussels. The compliance rate across all inspection
stations was estimated to be $80\%$ \citep{fischer_hybrid_2019}.
To find potentially suitable locations for inspections, we identified
pullouts across BC. We reduced the number of possible options by disregarding
some pullouts in close proximity to others. In total, we considered
$249$ location candidates.

Due to the large number of location candidates, we did not conduct
a detailed evaluation of the operational suitability of all considered
locations (e.g. pullout size, signage, and safety). Instead, we consulted
with the BC Invasive Mussel Defence Program to gauge the general suitability
of the locations suggested by the optimizer. If a suggested location
seemed unsuitable, we removed it from the candidate set and repeated
the optimization procedure. Despite this superficial suitability check,
a more detailed analysis would be necessary to account for all potential
practical constraints. These must be considered independent of the
model before an inspection station can be placed.

For each location, we assumed that $8\,\mt h$ long inspection shifts
could be started at each full hour of the day. Note that ``shift''
here refers to the time inspections are conducted and does not include
time required for staff to access or set up an inspection station.
The work time of staff will therefore be longer in practice. The assumed
length of the inspection shifts aligns with average operation patterns
of watercraft inspection stations in BC and divides each day in three
equally sized shifts, which simplifies the model. Though the effective
operation time (limited by access time of staff) is lower at remote
locations, our time model provides a good first approximation. 

\subsubsection{Cost model}

We determined the inspection costs based on correspondence with the
BC Invasive Mussel Defence Program. The considered optimization problem
is often easier to solve if costs are rounded to well aligned cost
units. Therefore, we set the infrastructure costs for setting up an
inspection station as our base cost unit. The costs per conducted
inspection shift are then $3.5$ units during daytime hours and $5.5$
units between $9\,\mt{PM}$ and $5\,\mt{AM}$. These costs include
salary, training, and equipment for inspection staff. In $2017$,
the BC Invasive Mussel Defence Program was operating on a budget of
approximately $80$ cost units.

As in-depth location-specific cost analysis would have been difficult,
we assumed that the inspection costs are equal for all considered
locations. Note, however, that site specific costs can vary significantly
and may be a limitation when assessing a location for overnight operations.

\subsection{Implementation}

As we considered about $300,000$ origin-destination pairs connected
by $6.7$ routes on average, considering all boater pathways individually
would be difficult. Therefore, we merged traffic of boaters passing
the same sets of potential inspection locations. As a result, the
number of distinct boater flows reduced to $2026$.

We determined the optimal inspection locations and operating times
under different budget scenarios. This allowed us to determine the
budget required to minimize the fraction of uninspected high-risk
boaters to a desired level. We also varied the model's noise term
to test how inspection strategies change under increased uncertainty.
To see how new infestations in close-by jurisdictions change the inspection
policy, we furthermore considered a scenario in which the American
states Idaho, Wyoming, and Oregon are invaded.

We implemented the model in the high-level programming language Python
version 3.7. To formulate the linear integer problem, we used the
modelling software CVXPY version 1.0.25 with added support for initial
guesses. To solve the linear integer problem, we used the commercial
solver MOSEK. We computed initial guesses with the greedy rounding
procedure described in section \ref{subsec:Solving-the-optimization-problems}.
We let the solver terminate if a solution with guaranteed accuracy
of $99.5\%$ was found or if $50$ minutes had passed. We conducted
the computations on a Linux server with a 20 core Intel Xeon 640 E5-2689
CPU (3.1GHz per CPU) and with 512GB RAM. The computer code can be
retrieved from \href{https://vemomoto.github.io/}{vemomoto.github.io}.

\section{Results}

In $72\%$ of the considered scenarios, we were able to identify a
solution with the desired accuracy of $99.5\%$. In the remaining
cases, the guaranteed solution quality never fell below $92\%$;
in scenarios with budgets $B\geq25$ units, we could always identify
solutions with $98\%$ accuracy and above. The greedy algorithm used
to compute an initial guess provided a solution with $99.5\%$ accuracy
in $58\%$ of the considered cases. The initial guesses always had
a quality above $90\%$.

\begin{figure}
\begin{centering}
\includegraphics[width=1\textwidth]{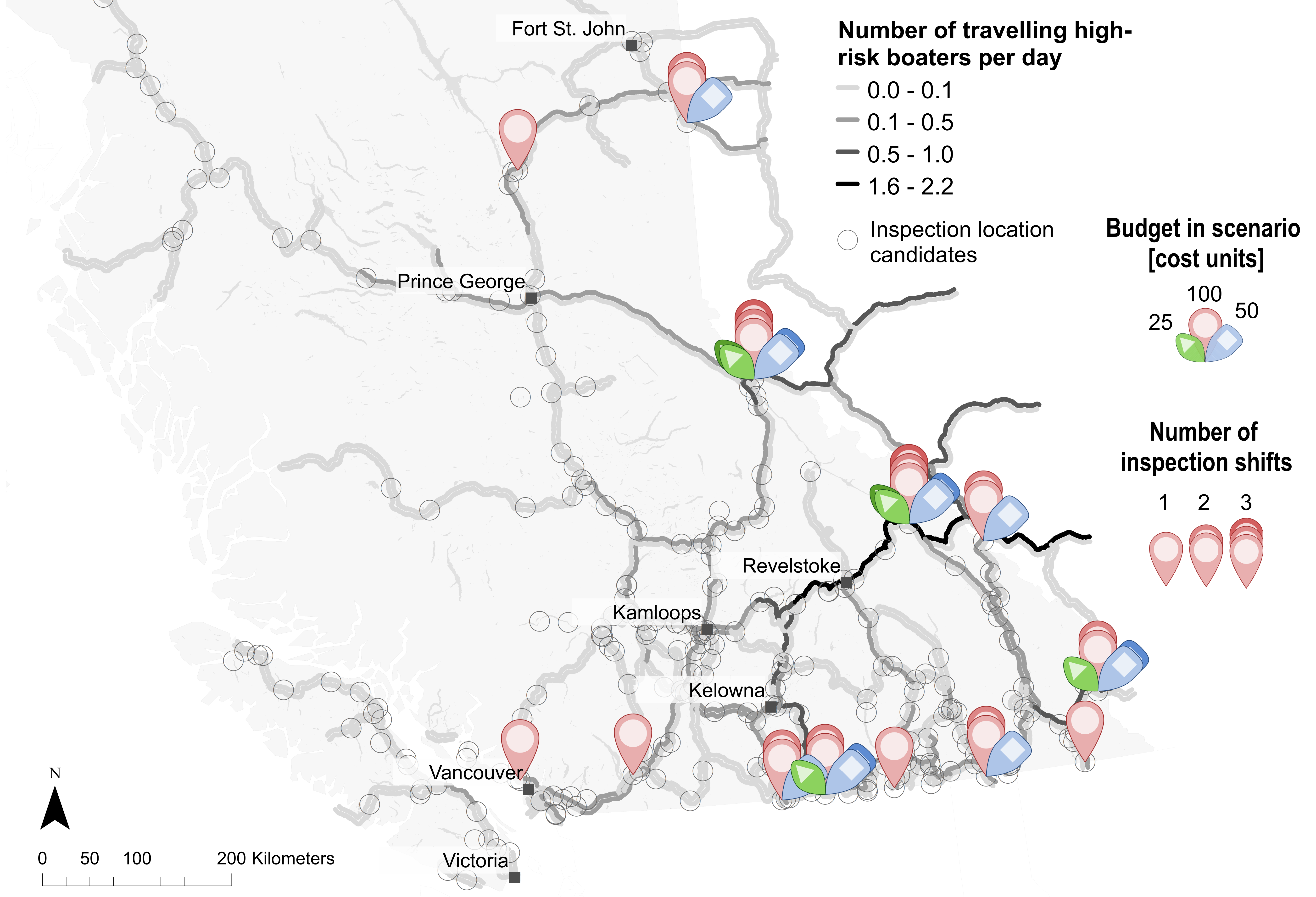}
\par\end{centering}
\caption[Optimal locations and operation shifts for three different budget
scenarios]{Optimal locations and operation shifts for three different budget
scenarios. Most inspection stations are placed close to the British
Columbian border. The markers depict the optimal inspection locations
for each scenario. Green (triangle): optimal locations with a budget
of $25$ units; blue (square) $50$ unit budget; red (circle) $100$
unit budget. The number of markers stacked on top of each other corresponds
to the optimal numbers of inspection shifts. The darkness of the roads
show the estimated boater traffic volume. The hollow circles depict
the considered candidates for inspection locations. \label{fig:Optimal-locations}}
\end{figure}

Figure \ref{fig:Optimal-locations} displays the optimized locations
and operating times for watercraft inspection stations in the considered
model scenario. We depict the respective optimal policy under three
different budget constraints. The optimal locations for inspections
are located close to border crossings if suitable locations are available.
However, where the traffic through many border crossings merges on
a major highway (e.g. in the Vancouver metropolitan area), it is optimal
to place the inspection stations farther inland.

\begin{figure}
\captionsetup[subfigure]{position=top,justification=raggedleft,margin=0pt, singlelinecheck=off}

\subfloat[\label{fig:characteristics-budget}]{\includegraphics[width=0.48\textwidth]{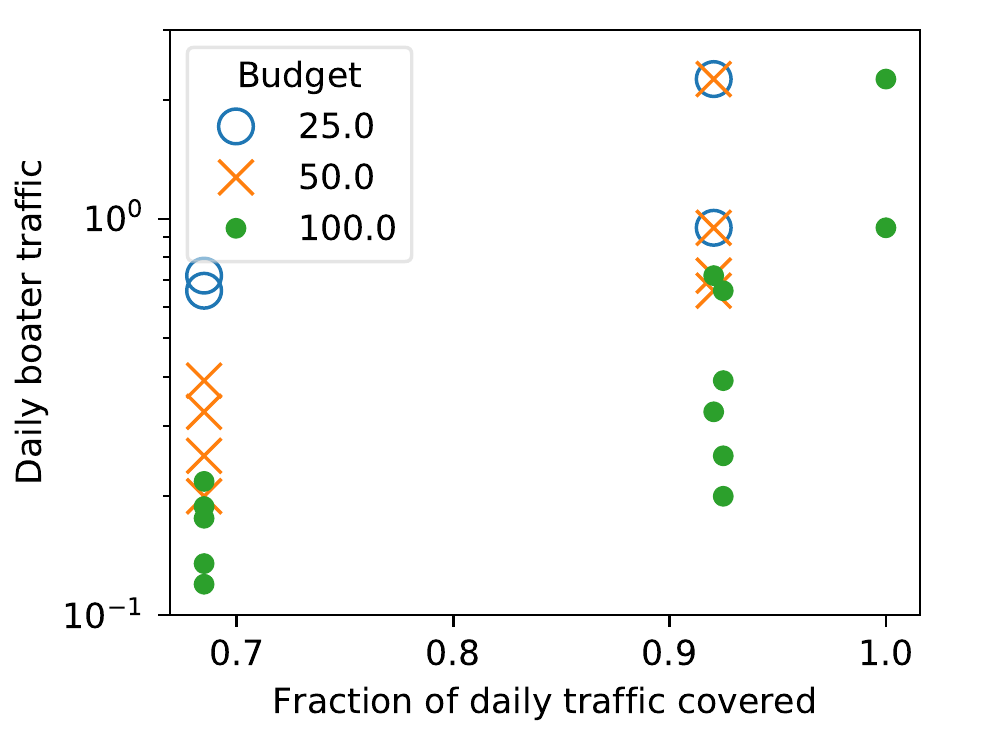}

}\hspace*{\fill}\subfloat[\label{fig:characteristics-noise}]{\includegraphics[width=0.48\textwidth]{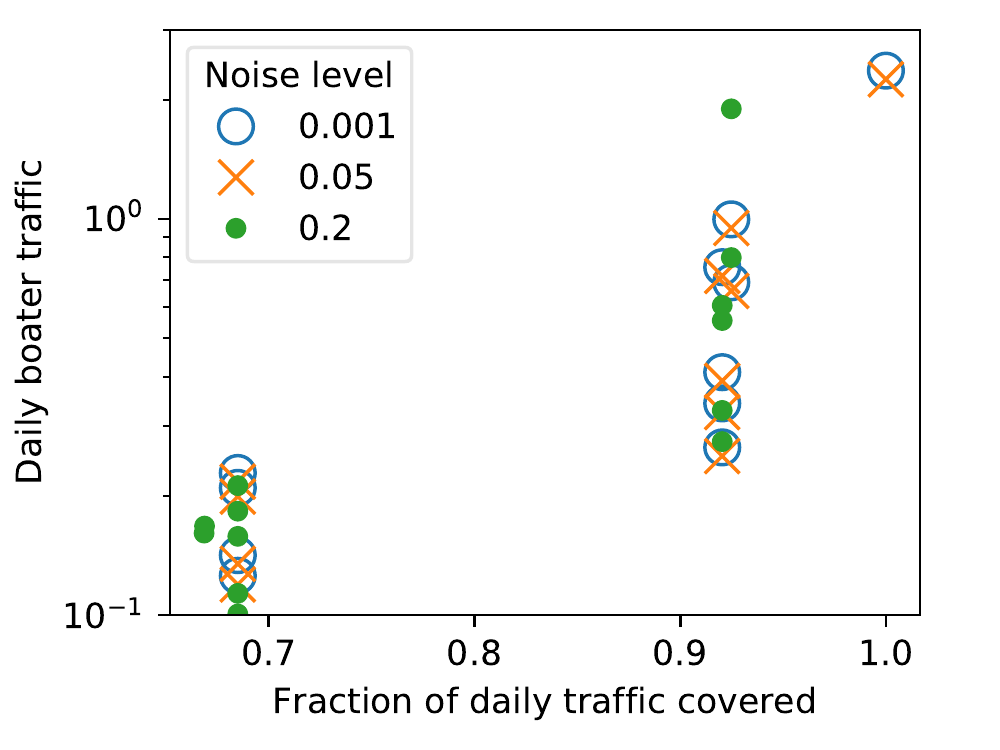}

}

\caption[Characteristics of the optimized inspection stations in different
scenarios.]{Characteristics of the optimized inspection stations in scenarios
with (a) different budget constraints, and (b) different levels of
uncertainty. Additional budget is preferably spent on additional inspection
locations rather than longer operating hours. Increased uncertainty
results in resources being distributed over more locations at cost
of decreasing operating hours. Overall, however, uncertainty does
not have a strong effect on the inspection policy. Each marker corresponds
to an inspection station. The position of a marker depicts the daily
traffic volume expected at the location and the fraction of daily
traffic covered under the optimal operation policy (compliance supposed).
The ``noise level'' denotes the fraction $\eta_{c}$ of boaters
travelling on routes not covered by the route choice model. Note that
the noise level also affects the daily traffic volume at the inspection
locations.\label{fig:Characteristics}}
\end{figure}

Figure \ref{fig:Characteristics} depicts characteristics of the optimal
inspection stations in different scenarios. The expected traffic volume
at an inspection station coincides with the optimized operating times:
stations with high expected boater traffic are operated longer than
stations with lower traffic. If the budget is increased, some stations
are assigned longer operating times. However, larger portions of the
additional budget are spent on additional locations (see also Figure
\ref{fig:Optimal-locations}). If the uncertainty in the traffic predictions
increases, more inspection stations are set up at the cost of shorter
operations. Overall, however, the noise level has little effect on
the inspection policy.

\begin{figure}[t]
\captionsetup[subfigure]{position=top,justification=raggedleft,margin=0pt, singlelinecheck=off}

\subfloat[\label{fig:covered-boater-fraction}]{\includegraphics[width=0.48\textwidth]{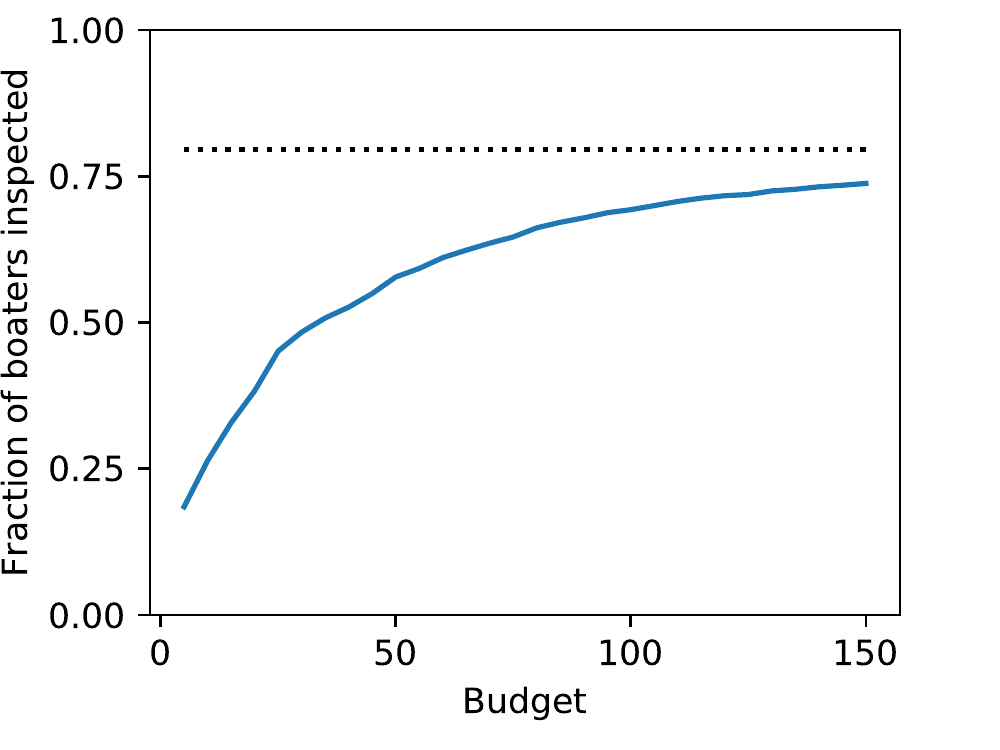}

}\hspace*{\fill}\subfloat[\label{fig:boater-price}]{\includegraphics[width=0.48\textwidth]{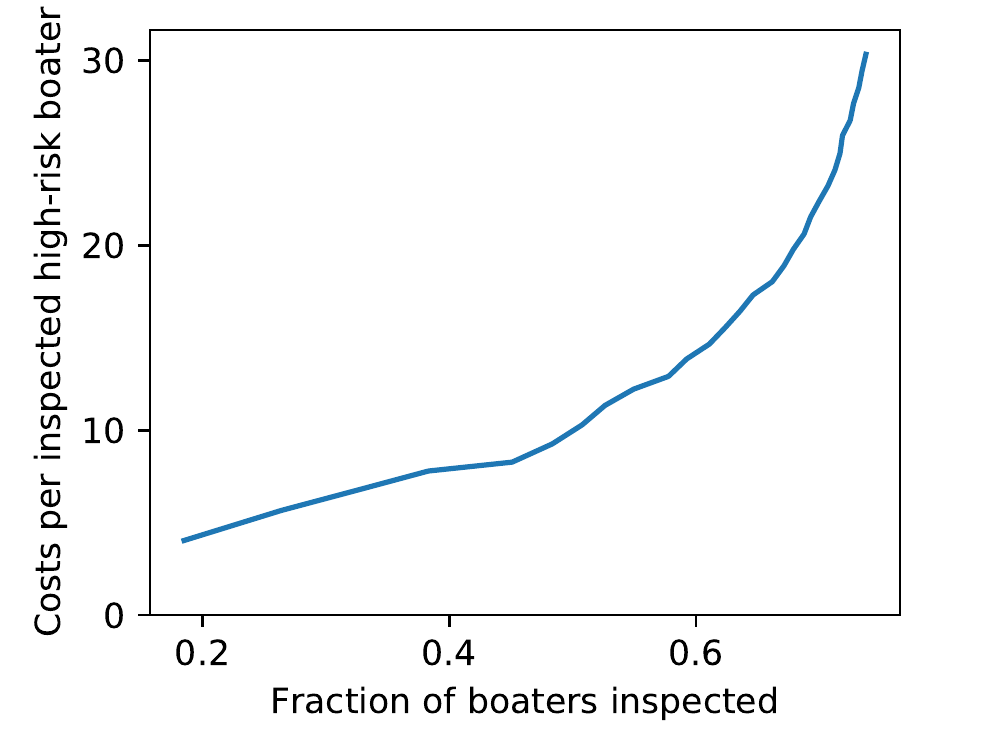}

}

\caption[Inspection effectiveness dependent on the budget constraint and price
per inspected high-risk boater dependent on the proportion of inspected
boaters]{ Inspection effectiveness dependent on the budget constraint (a)
and price per inspected high-risk boater dependent on the proportion
of inspected boaters (b). While a large fraction of high-risk boaters
can be covered with moderate effort, inspecting all complying boaters
is costly. Panel (a) shows the expected fraction of incoming high-risk
boaters that can be inspected under the optimal policy. The dotted
line shows the level of complying boaters, which is the maximal fraction
of boaters that can be inspected. \label{fig:budget-optimization}}
\end{figure}

Optimizing inspection station operation under a range of different
budget allowances showed that a moderate inspection budget, corresponding
to about half the $2017$ BC inspection budget, suffices to inspect
half of the incoming high-risk boaters (Figure \ref{fig:budget-optimization}).
However, the resources required for inspections increase quickly if
more boaters shall be controlled. Thereby, the faction of inspected
boaters is limited by boaters' compliance with inspections.

The considered change in the invasion state of three American states
had only a moderate impact on inspection policy. The results are depicted
in Supplementary Appendix B. As the additional propagule sources were
located south of BC, the inspection effort increased at the southern
border under the optimal policy. Furthermore, the optimal policy contained
less overnight inspections and distributed resources more evenly across
inspection stations. 

\section{Discussion}

We presented a method to optimize placement and operating times of
watercraft inspection stations. The approach is suited to model management
scenarios on a detailed level and gives specific advice for management
actions. We applied our approach to invasive mussel management in
BC and investigated the impact of budget constraints, model uncertainty,
and potential future invasions on management actions and efficiency.
However, it must be recognized that our model did not account for
all critical operational factors, such as site safety. Nonetheless,
the presented results provide valuable insights into optimal management
of AIS when combined with critical operational factors.

Most of our results are consistent with common sense. In general,
it is optimal to inspect boaters as soon as they enter the managed
region. That way, waterbodies close to the border can be protected.
If multiple routes via different border crossings merge close to the
border, it can be optimal to inspect boaters after this merging point.
Inspections stations should operate longer at locations with high
traffic volume. Furthermore, uncertainty in traffic predictions increases
the benefit of spreading the inspection efforts over many locations.
Driven by these simple principles, our results were remarkably robust
throughout considered scenarios and agree well with the watercraft
inspection policy currently implemented in BC.

While these qualitative principles may seem obvious, it can be challenging
to identify quantitative definitions of terms like ``close to the
border'' and ``longer''. The difficulty in optimizing management
policies is in balancing trade-offs, such as between leaving some
waterbodies close to the border unprotected and maximizing the overall
number of inspected boaters, or between long-time operation of few
highly frequented inspection stations and distribution of resources
over many locations. As the approach proposed in this paper is suited
to account for these trade-offs, it is a valuable extension to earlier
more theoretical results on AIS management \citep{potapov_allee_2008,potapov_optimal_2008,finnoff_control_2010}.

Considering scenarios with different budget constraints allowed us
to investigate the trade-off between resources invested in AIS control
and the number of inspected high-risk boaters. In combination with
the expected monetary damage caused by the arrival of an uncontrolled
boater at an uninvaded lake, this trade-off curve can be used to identify
the optimal budget for inspections. Since both invasion risk and damages
due to invasions are difficult to quantify, a rigorous computation
of the optimal inspection budget may not always be feasible in practice.
Nonetheless, the cost-effectiveness curve provides an estimate of
the efficacy of control efforts and shows which budget is required
to achieve a certain management goal.

In the case of AIS control in BC, a moderate budget suffices to inspect
a significant portion of the incoming high-risk boaters. This is because
boater traffic in BC concentrates on a small number of major highways.
Nevertheless, an attempt to inspect all high-risk boaters would be
very costly, as many minor roads would have to be considered as well.
It could therefore be more cost-effective to implement measures to
increase the compliance of boaters, e.g. through additional road signs
or public outreach and education.

We see particular use of our approach in its potential to optimize
rapid response actions under scenarios of interest. The extended invasion
scenario considered in this paper shows that slight adjustments to
the inspection policy may suffice to react on the new conditions.
In a similar manner, our approach could be used to assess the benefit
from cross-border collaborations, in which inspection efforts are
combined to control the boater inflow to a large joint area. Due to
the flexibility of our model, managers can consider a variety of scenarios
at little cost.

\subsection{Limitations and possible extensions}

The accuracy of our approach in real-world applications is strongly
dependent on the accuracy and level of detail of the utilized data
and models. Therefore, the results should be combined with expert
knowledge and refined iteratively if necessary. Nonetheless, our approach
can be extended to account for many management constraints and is
thus a helpful tool to optimize inspection policies.

Limitations exist with respect to the considered objective function.
Though the number of potentially infested watercraft arriving at a
waterbody is a valuable proxy for invasion risk, the establishment
probability of dreissenid mussels is not linear in propagule pressure
\citep{leung_predicting_2004}. Hence, our approach is not suited
to minimize invasion risk directly. However, high-dimensional non-convex
optimization problems are difficult to solve, and minimizing a proxy
for invasion risk may thus be the better option in practice. Nonetheless,
significant realism could be added by considering the suitability
of the destination waterbodies as habitat for AIS. This could be done
by weighting boater flows differently dependent on the invasion risk
of the destination waterbodies.

Since our traffic model does not explicitly account for the time boaters
need to travel between locations, the optimized inspection station
operating times may have to be adjusted to local temporal traffic
patterns. Though this shortage in model realism could affect the results
significantly if boaters pass multiple inspection stations under the
optimal policy, optimal inspection locations are often on independent
routes. In the scenario considered in this study, the optimized operating
times were all centered around the traffic peak. This indicates that
interactions between locations did not affect the operating times
and the error due to the simplifying model assumption is small.

Another modelling challenge is to account for uncertainty appropriately.
The noise model used in this study is a non-informative null model
that treats all potential inspection locations equally. As more boater
traffic may be expected at major highways than at minor roads, the
noise model could be improved by incorporating location-dependent
covariates. However, since our results were not very sensitive to
the noise level, a realistic noise model might not change the optimal
policy significantly.

Our model did not incorporate site-specific costs and operational
constraints. In high-budget scenarios, this let our model suggest
overnight inspections at remote sites that are lacking the required
infrastructure to safely operate at night. Requirements for overnight
inspections include proper road infrastructure (lanes/barriers), lighting,
access to safe communication and nearby living accommodations for
staff. A lack of living accommodations for staff can also limit the
number of staff based in remote locations. These constraints could
be incorporated in a more detailed model as well as increased costs
at remote locations. A more detailed model could also account for
inspection stations operated by neighbouring jurisdictions. As an
example, the BC program works closely with the Canadian Border Services
Agency and neighbouring provinces and states to receive advanced notifications
of high risk watercraft destined for BC. Nonetheless, the presented
model includes major factors affecting inspection station operation.
Therefore, the model can serve as a helpful resource to inform managers'
decisions in parallel with operational constraints.

Another potential extension of our model is to incorporate location-specific
or management-dependent compliance rates. At certain sites, such as
cross-national border crossings, compliance can be enforced more easily
than at other locations. Compliance may furthermore depend on management
efforts: it may be possible to increase the compliance rate of boaters
at some costs. In Supplementary Appendix C, we show how non-uniform
and flexible compliance rates can be considered with small model adjustments.

The computational method we used to optimize inspection station operation
is well established and builds on a large body of theoretical insights
\citep{ageev_pipage_2004,conforti_integer_2014}. Nonetheless, the
problem is computational difficult, and there may be scenarios in
which linear integer solvers fail to provide good solutions. Optimization
failures are most prevalent in scenarios in which portions of the
budget remain unused under the optimal policy or in which many boaters
pass multiple inspection stations under optimized operation. In both
cases, the solution to the continuous relaxation of the problem may
differ significantly from the integer solution.

However, issues due to unused budget become minor if the considered
budget is sufficiently large. Furthermore, the issue may be mitigated
by adjusting the budget slightly. Issues with redundant inspection
stations, in turn, are unlikely to occur if the propagule donors and
recipients are in separate regions. Then, independent inspection locations
can often be identified. This is often the case if invasion processes
are considered on large scales. Therefore, our approach will yield
good results in most applications. We provide more details in Supplementary
Appendix D.

\subsection{General conclusions for invasive species management}

In this paper, we considered specific management scenarios with focus
of AIS control in BC. Nonetheless, some common patterns were consistent
throughout our results and may thus apply with greater generality.
These principles may be used as rules of thumb if no comprehensive
modelling and optimization effort is possible. Below we summarize
these conclusions.
\begin{itemize}
\item Inspection stations should be placed close to the border of the uninfested
region. Consequently, cross-border collaborations between uninvaded
jurisdictions have a high potential of improving the cost-effectiveness
of control.
\item If traffic flows merge close to the border, inspections are more cost-effective
after the merging point. Hence, identifying such points is crucial
for successful management.
\item If traffic predictions involve a high level of uncertainty, inspection
efforts should be distributed over many locations at the cost lower
inspection effort at each site.
\item If a high reduction of the propagule inflow is desired, it may be
most cost-effective to implement measures increasing the compliance
rate rather than operating more inspection stations for longer hours.
\end{itemize}

\section*{Authors' contributions}

All authors conceived the project; SMF conceived the methods jointly
with MAL. SMF and MB jointly prepared the data for the analysis. SMF
conducted the mathematical analysis, implemented the model, and wrote
the manuscript. All authors revised the manuscript.

\section*{Acknowledgements}

The authors would like to give thanks to the BC Ministry of Environment
and Climate Change Strategy staff of the BC Invasive Mussel Defence
Program, who conducted the survey this study is based on. Furthermore,
the authors thank the members of the Lewis Research Group at the University
of Alberta for helpful feedback and discussions. SMF is thankful for
the funding received from the Canadian Aquatic Invasive Species Network
and the Natural Sciences and Engineering Research Council of Canada
(NSERC); MAL gratefully acknowledges an NSERC Discovery Grant and
Canada Research Chair.

\section*{Competing Interests}

The authors declare no competing interests.

\section*{}

\bibliographystyle{apalike}

\begin{thebibliography}{}

\bibitem[Ageev and Sviridenko, 2004]{ageev_pipage_2004}
Ageev, A. and Sviridenko, M. (2004).
\newblock Pipage rounding: a new method of constructing algorithms with proven
  performance guarantee.
\newblock {\em Journal of Combinatorial Optimization}, 8(3):307--328.

\bibitem[Ageev and Sviridenko, 1999]{cornuejols_approximation_1999}
Ageev, A.~A. and Sviridenko, M.~I. (1999).
\newblock Approximation algorithms for maximum coverage and max cut with given
  sizes of parts.
\newblock In Cornu{\'e}jols, G., Burkard, R.~E., and Woeginger, G.~J., editors,
  {\em Integer {Programming} and {Combinatorial} {Optimization}}, volume 1610,
  pages 17--30. Springer Berlin Heidelberg, Berlin, Heidelberg.

\bibitem[{Alberta Environment and Parks Fish and Wildlife Policy},
  2015]{alberta_environment_and_parks_fish_and_wildlife_policy_alberta_2015}
{Alberta Environment and Parks Fish and Wildlife Policy} (2015).
\newblock Alberta {Aquatic} {Invasive} {Species} {Program} 2015 annual report.
\newblock Technical report, Edmonton, AB.

\bibitem[Bovy, 2009]{bovy_modelling_2009}
Bovy, P. H.~L. (2009).
\newblock On modelling route choice sets in transportation networks: a
  synthesis.
\newblock {\em Transport Reviews}, 29(1):43--68.

\bibitem[Conforti et~al., 2014]{conforti_integer_2014}
Conforti, M., Cornu{\'e}jols, G., and Zambelli, G. (2014).
\newblock {\em Integer programming}, volume 271 of {\em Graduate {Texts} in
  {Mathematics}}.
\newblock Springer International Publishing, Cham.

\bibitem[Epanchin-Niell and Wilen, 2012]{epanchin-niell_optimal_2012}
Epanchin-Niell, R.~S. and Wilen, J.~E. (2012).
\newblock Optimal spatial control of biological invasions.
\newblock {\em Journal of Environmental Economics and Management},
  63(2):260--270.

\bibitem[Feige, 1998]{feige_threshold_1998}
Feige, U. (1998).
\newblock A threshold of ln n for approximating set cover.
\newblock {\em Journal of the ACM}, 45(4):634--652.

\bibitem[Finnoff et~al., 2010]{finnoff_control_2010}
Finnoff, D., Potapov, A., and Lewis, M.~A. (2010).
\newblock Control and the management of a spreading invader.
\newblock {\em Resource and Energy Economics}, 32(4):534--550.

\bibitem[Fischer, 2019]{fischer_locally_2019}
Fischer, S.~M. (2019).
\newblock Locally optimal routes for route choice sets.
\newblock {\em arXiv e-prints}, pages 1--40.
\newblock arXiv:1909.08801.

\bibitem[Fischer et~al., 2019]{fischer_hybrid_2019}
Fischer, S.~M., Beck, M., Herborg, L.-M., and Lewis, M.~A. (2019).
\newblock A hybrid gravity and route choice model to assess vector traffic in
  large-scale road networks.
\newblock {\em arXiv e-prints}, pages 1--26.
\newblock arXiv:1909.08811.

\bibitem[Hastings et~al., 2006]{hastings_simple_2006}
Hastings, A., Hall, R.~J., and Taylor, C.~M. (2006).
\newblock A simple approach to optimal control of invasive species.
\newblock {\em Theoretical Population Biology}, 70(4):431--435.

\bibitem[Hyyti{\"a}inen et~al., 2013]{hyytiainen_optimization_2013}
Hyyti{\"a}inen, K., Lehtiniemi, M., Niemi, J.~K., and Tikka, K. (2013).
\newblock An optimization framework for addressing aquatic invasive species.
\newblock {\em Ecological Economics}, 91:69--79.

\bibitem[{Inter-Ministry Invasive Species Working Group},
  2015]{inter-ministry_invasive_species_working_group_zebra_2015}
{Inter-Ministry Invasive Species Working Group} (2015).
\newblock Zebra and quagga mussel early detection and rapid response plan.
\newblock Technical report, Victoria, BC.

\bibitem[Johnson et~al., 2001]{johnson_overland_2001}
Johnson, L.~E., Ricciardi, A., and Carlton, J.~T. (2001).
\newblock Overland dispersal of aquatic invasive species: a risk assessment of
  transient recreational boating.
\newblock {\em Ecological Applications}, 11(6):1789--1799.

\bibitem[Johnson et~al., 2017]{johnson_invasive_2017}
Johnson, R., Crafton, R.~E., and Upton, H.~F. (2017).
\newblock Invasive species: major laws and the role of selected federal
  agencies.
\newblock Technical report, Congressional Research Service, Washington, DC.

\bibitem[Karatayev et~al., 2015]{karatayev_zebra_2015}
Karatayev, A.~Y., Burlakova, L.~E., and Padilla, D.~K. (2015).
\newblock Zebra versus quagga mussels: a review of their spread, population
  dynamics, and ecosystem impacts.
\newblock {\em Hydrobiologia}, 746(1):97--112.

\bibitem[Khuller et~al., 1999]{khuller_budgeted_1999}
Khuller, S., Moss, A., and Naor, J.~S. (1999).
\newblock The budgeted maximum coverage problem.
\newblock {\em Information Processing Letters}, 70(1):39--45.

\bibitem[K{\i}b{\i}{\c s} and B{\"u}y{\"u}ktahtak{\i}n,
  2017]{kibis_optimizing_2017}
K{\i}b{\i}{\c s}, E.~Y. and B{\"u}y{\"u}ktahtak{\i}n, {\.I}.~E. (2017).
\newblock Optimizing invasive species management: a mixed-integer linear
  programming approach.
\newblock {\em European Journal of Operational Research}, 259(1):308--321.

\bibitem[Lee, 2010]{lee_circular_2010}
Lee, A. (2010).
\newblock Circular data.
\newblock {\em Wiley Interdisciplinary Reviews: Computational Statistics},
  2(4):477--486.

\bibitem[Leung et~al., 2004]{leung_predicting_2004}
Leung, B., Drake, J.~M., and Lodge, D.~M. (2004).
\newblock Predicting invasions: {Propagule} pressure and the gravity of {Allee}
  effects.
\newblock {\em Ecology}, 85(6):1651--1660.

\bibitem[Leung et~al., 2002]{leung_ounce_2002}
Leung, B., Lodge, D.~M., Finnoff, D., Shogren, J.~F., Lewis, M.~A., and
  Lamberti, G. (2002).
\newblock An ounce of prevention or a pound of cure: bioeconomic risk analysis
  of invasive species.
\newblock {\em Proceedings of the Royal Society B: Biological Sciences},
  269(1508):2407--2413.

\bibitem[Lockwood et~al., 2013]{lockwood_invasion_2013}
Lockwood, J.~L., Hoopes, M.~F., and Marchetti, M.~P. (2013).
\newblock {\em Invasion ecology}.
\newblock Wiley-Blackwell, Chichester, West Sussex, UK, 2nd edition.

\bibitem[Mangin, 2011]{mangin_100th_2011}
Mangin, S. (2011).
\newblock The 100th {Meridian} {Initiative}: a strategic approach to prevent
  the westward spread of zebra mussels and other aquatic nuisance species.
\newblock Technical Report 152, U.S. Fish and Wildlife Service, Arlington, VA.

\bibitem[Pimentel et~al., 2005]{pimentel_update_2005}
Pimentel, D., Zuniga, R., and Morrison, D. (2005).
\newblock Update on the environmental and economic costs associated with
  alien-invasive species in the {United} {States}.
\newblock {\em Ecological Economics}, 52(3):273--288.

\bibitem[Potapov and Lewis, 2008]{potapov_allee_2008}
Potapov, A.~B. and Lewis, M.~A. (2008).
\newblock Allee effect and control of lake system invasion.
\newblock {\em Bulletin of Mathematical Biology}, 70(5):1371--1397.

\bibitem[Potapov et~al., 2008]{potapov_optimal_2008}
Potapov, A.~B., Lewis, M.~A., and Finnoff, D.~C. (2008).
\newblock Optimal control of biological invasions in lake networks.
\newblock {\em Natural Resource Modeling}, 20(3):351--379.

\bibitem[Rosaen et~al., 2012]{rosaen_costs_2012}
Rosaen, A.~L., Grover, E.~A., and Spencer, C.~W. (2012).
\newblock The costs of aquatic invasive species to {Great} {Lakes} states.
\newblock Technical report, Anderson Economical Group, East Lansing, MI.

\bibitem[Shine et~al., 2010]{shine_assessment_2010}
Shine, C., Kettunen, M., Genovesi, P., Essl, F., Gollasch, S., Rabitsch, W.,
  Scalera, R., Starfinger, U., and ten Brink, P. (2010).
\newblock Assessment to support continued development of the {EU} {Strategy} to
  combat invasive alien species.
\newblock Final {Report} for the {European} {Commission}, Institute for
  European Environmental Policy (IEEP), Brussels, Belgium.

\bibitem[Surkov et~al., 2008]{surkov_model_2008}
Surkov, I.~V., Oude~Lansink, A. G. J.~M., van Kooten, O., and van~der Werf, W.
  (2008).
\newblock A model of optimal import phytosanitary inspection under capacity
  constraint.
\newblock {\em Agricultural Economics}, 38(3):363--373.

\bibitem[Trullols et~al., 2010]{trullols_planning_2010}
Trullols, O., Fiore, M., Casetti, C., Chiasserini, C., and Barcelo~Ordinas, J.
  (2010).
\newblock Planning roadside infrastructure for information dissemination in
  intelligent transportation systems.
\newblock {\em Computer Communications}, 33(4):432--442.

\bibitem[Turbelin et~al., 2017]{turbelin_mapping_2017}
Turbelin, A.~J., Malamud, B.~D., and Francis, R.~A. (2017).
\newblock Mapping the global state of invasive alien species: patterns of
  invasion and policy responses.
\newblock {\em Global Ecology and Biogeography}, 26(1):78--92.

\bibitem[Vander~Zanden and Olden, 2008]{vander_zanden_management_2008}
Vander~Zanden, M.~J. and Olden, J.~D. (2008).
\newblock A management framework for preventing the secondary spread of aquatic
  invasive species.
\newblock {\em Canadian Journal of Fisheries and Aquatic Sciences},
  65(7):1512--1522.

\end{thebibliography}

\end{document}


\global\long\def\argparentheses#1{\mathopen{\left(#1\right)}\mathclose{}}%
\global\long\def\ap#1{\argparentheses{#1}\mathclose{}}%

\global\long\def\var#1{\mathbb{V}\argparentheses{#1}}%
\global\long\def\pr#1{\mathbb{P}\argparentheses{#1}}%
\global\long\def\ev#1{\mathbb{E}\argparentheses{#1}}%

\global\long\def\smft#1#2{\stackrel[#1]{#2}{\sum}}%
\global\long\def\smo#1{\underset{#1}{\sum}}%
\global\long\def\prft#1#2{\stackrel[#1]{#2}{\prod}}%
\global\long\def\pro#1{\underset{#1}{\prod}}%
\global\long\def\uno#1{\underset{#1}{\bigcup}}%

\global\long\def\order#1{\mathcal{O}\argparentheses{#1}}%
\global\long\def\R{\mathbb{R}}%
\global\long\def\Q{\mathbb{Q}}%
\global\long\def\N{\mathbb{N}}%
\global\long\def\Z{\mathbb{Z}}%
\global\long\def\F{\mathcal{F}}%

\global\long\def\mathtext#1{\mathrm{#1}}%
\global\long\def\mt#1{\mathtext{#1}}%

\global\long\def\maxo#1{\underset{#1}{\max\,}}%
\global\long\def\argmaxo#1{\underset{#1}{\mathtext{argmax}\,}}%
\global\long\def\minargmaxo#1{\underset{#1}{\mathtext{minargmax}\,}}%
\global\long\def\argsupo#1{\underset{#1}{\mathtext{argsup}\,}}%
\global\long\def\supo#1{\underset{#1}{\sup\,}}%
\global\long\def\info#1{\underset{#1}{\inf\,}}%
\global\long\def\mino#1{\underset{#1}{\min\,}}%
\global\long\def\argmino#1{\underset{#1}{\mathtext{argmin}\,}}%
\global\long\def\limo#1#2{\underset{#1\rightarrow#2}{\lim}}%
\global\long\def\supo#1{\underset{#1}{\sup}}%
\global\long\def\info#1{\underset{#1}{\inf}}%

\global\long\def\b#1{\boldsymbol{#1}}%
\global\long\def\ol#1{\overline{#1}}%
\global\long\def\ul#1{\underline{#1}}%

\newcommandx\der[3][usedefault, addprefix=\global, 1=, 2=, 3=]{\frac{d^{#2}#3}{d#1^{#2}}}%
\newcommandx\pder[3][usedefault, addprefix=\global, 1=, 2=]{\frac{\partial^{#2}#3}{\partial#1^{#2}}}%
\global\long\def\intft#1#2#3#4{\int\limits _{#1}^{#2}#3d#4}%
\global\long\def\into#1#2#3{\underset{#1}{\int}#2d#3}%

\global\long\def\th{\theta}%
\global\long\def\la#1{~#1~}%
\global\long\def\laq{\la =}%
\global\long\def\normal#1#2{\mathcal{N}\argparentheses{#1,\,#2}}%
\global\long\def\uniform#1#2{\mathcal{U}\argparentheses{#1,\,#2}}%
\global\long\def\I#1#2{\mbox{I}_{#1}\argparentheses{#2}}%
\global\long\def\chisq#1{\chi_{#1}^{2}}%
\global\long\def\dar{\,\Longrightarrow\,}%
\global\long\def\dal{\,\Longleftarrow\,}%
\global\long\def\dad{\,\Longleftrightarrow\,}%
\global\long\def\norm#1{\left\Vert #1\right\Vert }%
\global\long\def\code#1{\mathtt{#1}}%
\global\long\def\descr#1#2{\underset{#2}{\underbrace{#1}}}%
\global\long\def\NB{\mathcal{NB}}%
\global\long\def\BNB{\mathcal{BNB}}%
\global\long\def\e#1{\text{e}_{#1}}%
\global\long\def\P{\mathbb{P}}%
\global\long\def\pb#1{\Bigg(#1\Bigg)}%
\global\long\def\cpb#1{\Big[#1\Big]}%

\global\long\def\vv#1{\boldsymbol{#1}}%

\global\long\def\True{\mathtt{True}}%

\global\long\def\False{\mathtt{False}}%

\title{Supplementary Appendices for ``Managing Aquatic Invasions: Optimal
Locations and Operating Times for Watercraft Inspection Stations''}

\maketitle

\appendix
\renewcommand{\theequation}{A\arabic{equation}}
\renewcommand{\thefigure}{A\arabic{figure}}
\renewcommand{\thetable}{A\arabic{table}}
\setcounter{figure}{0}
\setcounter{equation}{0}

\section{Greedy rounding algorithm\label{sec:Greedy-rounding-algorithm}}

In this Appendix, we describe the greedy rounding algorithm we applied
to obtain initial guesses for the general branch and bound solvers.
We start by introducing some helpful notation. Let $P$ be the linear
integer problem that we desire to solve and $P_{\mt{cont}}$ its continuous
relaxation, in which decision variables may attain fractional values.
We write $\vv x$ for the $N$-dimensional vector of decision variables,
indexed by $\left(l,s\right)\in L\times S$. Let $\vv e_{ls}$ be
a unit vector that is $0$ everywhere except for the component corresponding
to the index $\left(l,s\right)$. Suppose that $C(\text{\ensuremath{\vv x}})$
denotes the cost for implementing a policy given by $\vv x$. We provide
pseudo code for the greedy rounding algorithm in Algorithm \ref{alg:GreedyRounding}.

The algorithm repeatedly solves the relaxed problem $P_{\mt{cont}}$
with different constraints fixing some decision variables to integer
values. The algorithm proceeds in two phases. In the first phase,
the maximal non-integral decision variable that can be rounded up
without violating the budget constraint is determined. With this variable
fixed, problem $P_{\mt{cont}}$ is solved again. When no additional
component can be rounded up without violating the cost constraint,
all previous constraints are removed, and the set of utilized locations
is fixed instead. The algorithm sets a flag $locked$ to $\mathtt{True}$
to show that the second phase of the algorithm has started.

In the second phase, components of $\vv x$ are still rounded up if
possible. However, now we do not round up the largest non-integral
component of $\vv x$. Instead, we determine for some location $l\in L$
with non-integral operation (i.e. $\exists\tilde{s}\in S_{l}\,:\,x_{l\tilde{s}}\notin\left\{ 0,1\right\} $)
the first time interval 
\begin{eqnarray}
t & := & \minargmaxo{t\in T}\left\{ \smo{s\in S_{lt}}x_{ls}\Bigg|x_{ls}<1\forall s\in S_{lt}\right\} 
\end{eqnarray}
that is operated strongest at this location. Here, $\minargmaxo{}\left\{ \cdot\right\} $
refers to the minimal admissible value for $\argmaxo{}\left\{ \cdot\right\} $
if the maximum is not unique. Then, we round up the latest affordable
shift $s\in S_{l}$ that covers the time interval $t$ and add $x_{ls}=1$
to the set of constraints. If no additional shift can be operated
at location $l$, we add a constraint fixing the usage of this location:
$x_{ls}=\left\lfloor x_{ls}\right\rfloor $ for all $s\in S_{l}$.

Distinguishing between the two phases of the algorithm yields optimized
operating times. Suppose we are in phase $2$, and consider the example
depicted in Figure \ref{fig:Optimal-locations-Scenario}. The solution
to the relaxed problem $P_{\mt{cont}}$ suggests that $3$ inspection
shifts $s_{1}$, $s_{2}$, and $s_{3}$ are conducted fractionally
at the considered location $l$. Thereby, $s_{2}$ overlaps with $s_{1}$
and $s_{3}$. The respective operation intensities are $x_{ls_{1}}=x_{ls_{3}}=0.8$
and $x_{ls_{2}}=0.2$. The budget assigned to this location does not
suffice to operate both $s_{1}$ and $s_{3}$ completely. Hence, only
one shift can be operated at $l$. Naive greedy rounding would suggest
to operate shift $s_{1}$, as it is the earliest shift with the maximal
fractional operation. However, in the optimal solution, the time interval
between $8$ AM and $4$ PM should be operated strongest. Therefore,
shift $s_{2}$ would be the optimal choice.

In its second phase, the suggested algorithm rounds up shifts based
on the maximal \emph{cumulative} operation rather than choosing the
shift with the highest operation variable. Nonetheless, it would be
of disadvantage to apply this rounding scheme in phase $1$ of the
algorithm, in which the set of used locations is not fixed. In this
case, shifts in the middle of the day would always be chosen with
preference, which make operation of \emph{two} shifts on a day less
efficient. In the second phase, it is typically known how many shifts
should be operated at each location.

Slight improvements to the suggested algorithm are possible. For example,
we added constraints in phase $1$ to suppress fractional operation
of shifts that would not be affordable completely under the costs
of the already constrained variables. However, this improvement is
unlikely to have a major effect on the results.

\begin{algorithm}
\setstretch{1.35}
\SetKw{And}{and}
\SetKw{Not}{not}
\SetKw{Continue}{continue}
\SetKw{With}{with}
\SetKwFunction{LockLocation}{lock\_location}
\SetKwProg{Fn}{Function}{:}{}
\Fn{\LockLocation{$\tilde{\vv{x}}$, $l$, $\varTheta$}}{
	\ForEach{$s \in S_l$}{
		$\varTheta := \varTheta \cup \{x_{ls}=\tilde{x}_{ls}\}$\;
	}
}

$locked := \False$; $\varTheta := \emptyset$\;

\While{$\True$}{
	$\vv{x} := $ solution to $P_{\mt{cont}}$ subject to  additional constraints in $\varTheta$\;
	\If{$\vv{x}\in \Z^N$}{
		\Return $\vv{x}$\;
	}
	$\tilde{\vv{x}} := \left\lfloor \vv x\right\rfloor $\;
	$\varOmega:=\left\{ \left(l,s\right)\in L\times S\,|\,0<\vv x_{ls}<1;\,C(\tilde{\vv x}+\vv{e}_{ls})\leq B\right\} $\;
	\eIf{$\varOmega = \emptyset$}{
		\eIf{\Not $locked$}{
			$locked := \True$; $\varTheta := \emptyset$\;
			
			\ForEach{$l \in L$ \With $\maxo{s\in S_{l}}x_{ls}=1$}{
				$\varTheta := \varTheta \cup \{\maxo{s\in S_{l}}x_{ls}=1\}$\;
			}
		}{
			$l := $ some location with $0 < x_{ls} < 1$ for some $s\in S_l$\;
			\LockLocation{$\tilde{\vv{x}}$, $l$, $\varTheta$}\;
		}
	}{
		$\left(l,s\right):=\minargmaxo{\left(l,s\right)\in\varOmega}x_{ls}$\;
		\If{$locked$}{	
			$t:=\minargmaxo{t\in T}\left\{ \smo{s\in S_{lt}}x_{ls}\Bigg|x_{ls}<1\forall s\in S_{lt}\right\} $\;
			$\varPsi:=\left\{ s\in S_{lt}\big|C(\tilde{\vv x}+\vv{e}_{ls})\leq B\right\} $\;
			\eIf{$\varPsi = \emptyset$}{
				\LockLocation{$\tilde{\vv{x}}$, $l$, $\varTheta$}\;
				\Continue\;
			}{
				$s:=\max S_{lt}$\;
			}
		}
		$\varTheta := \varTheta \cup \{x_{ls}=1\}$\;
	}
}

\caption[Greedy rounding algorithm]{Greedy rounding algorithm.\label{alg:GreedyRounding}}
\end{algorithm}

\begin{figure}[t]
\begin{centering}
\includegraphics{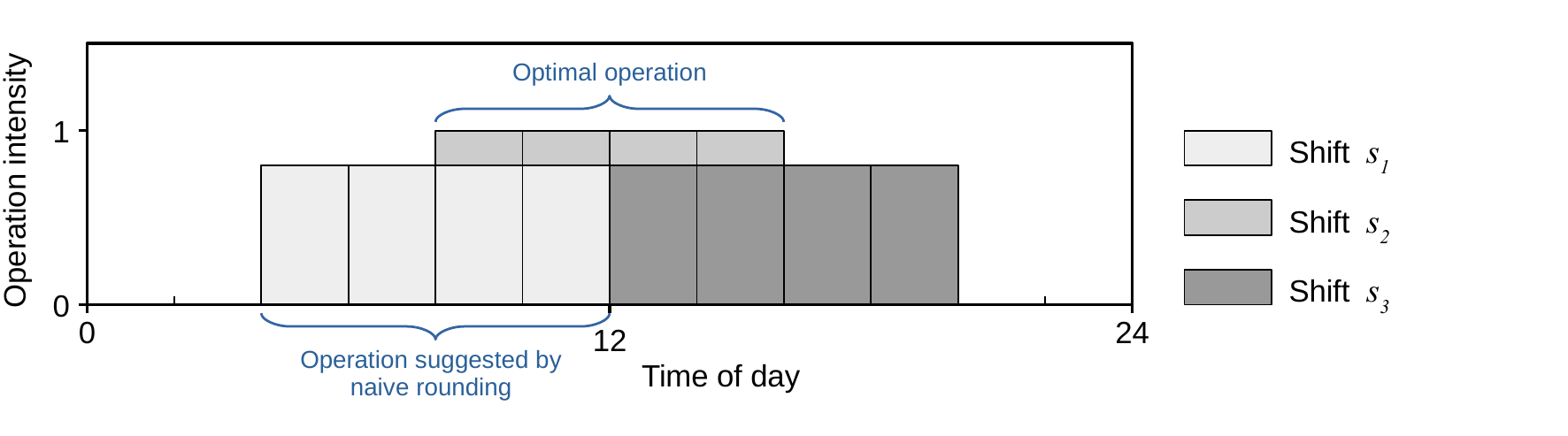}
\par\end{centering}
\caption[Motivation for the changed rounding procedure in phase $2$ of the
greedy rounding algorithm]{Motivation for the changed rounding procedure in phase $2$ of the
greedy rounding algorithm. The operation intensity is depicted as
a function of time for some inspection location. The intervals on
the time axis depict the discretization of the day time. The grey
boxes show the extent to which the inspection station would be operated
in the respective time intervals if fractional operation would be
allowed. The boxes' colours correspond to the respective operation
shifts. Naive greedy rounding would suggest to operate shift $s_{1}$.
Improved rounding, however, would prefer the time interval in which
the cumulative operation is maximal (shift $s_{2}$).}
\end{figure}

\section{Optimal inspection policy if additional parts of the USA are infested\label{sec:Optimal-inspection-policy-USA-infested}}

\begin{figure}
\begin{centering}
\includegraphics[width=1\textwidth]{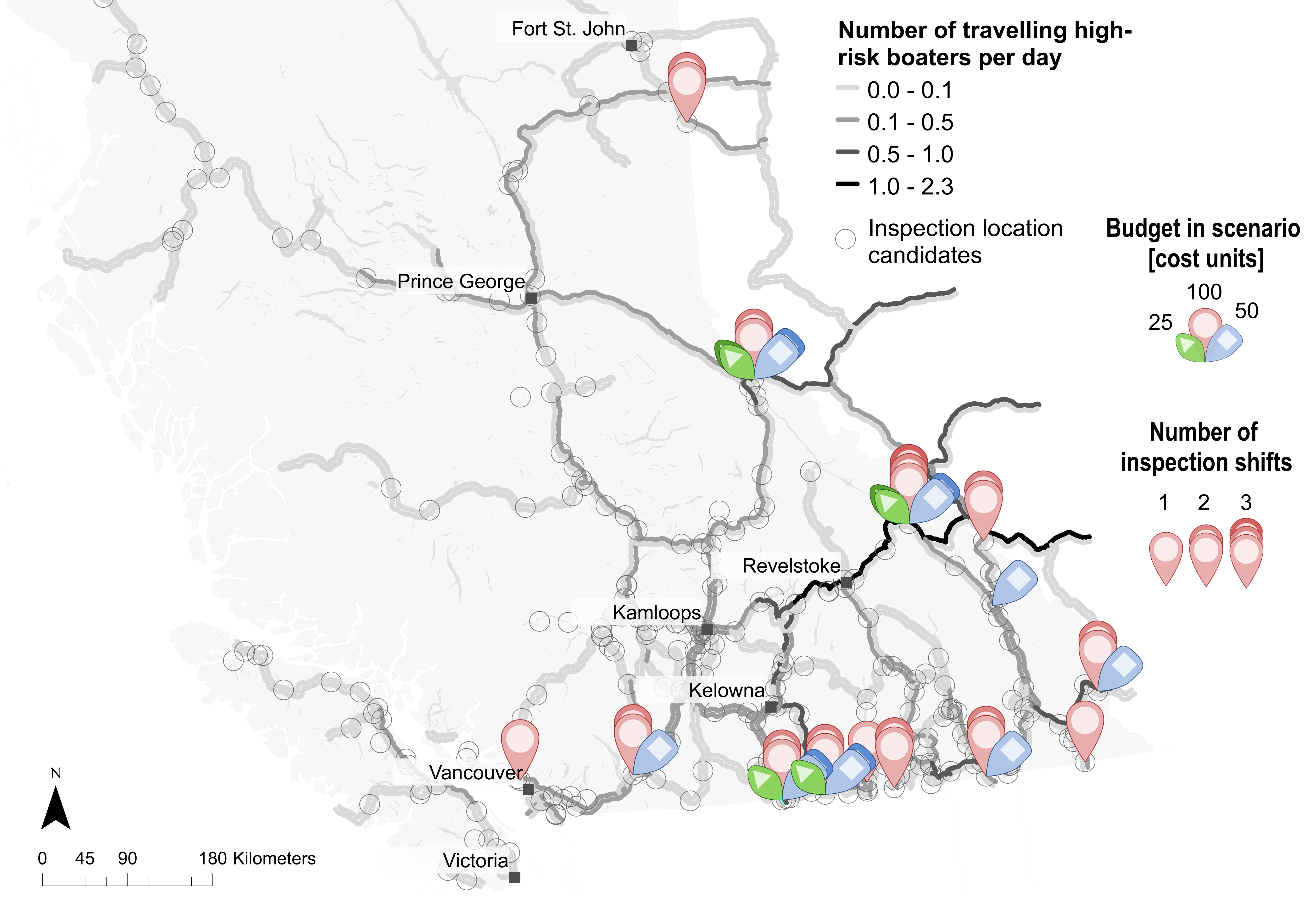}
\par\end{centering}
\caption[Optimal locations and operation shifts assuming that Idaho, Oregon,
and Wyoming are mussel invaded]{Optimal locations and operation shifts assuming that Idaho, Oregon,
and Wyoming are mussel invaded. Compared to the base scenario with
fewer infested States south of BC, more inspections are conducted
at the southern border of the province. The symbols have the same
meaning as in Figure 2 (main text). \label{fig:Optimal-locations-Scenario}}
\end{figure}

To assess how the optimal inspection policy changes if additional
states are infested, we considered a scenario in which boaters from
Idaho, Oregon, and Wyoming were considered high-risk boaters. The
results are depicted in Figure \ref{fig:Optimal-locations-Scenario}.
As more high-risk boaters enter BC via the southern border, inspection
efforts at this border are increased. The required resources are freed
by operating fewer inspection stations over night and by abandoning
inspection locations in the north. Nonetheless, the overall changes
are moderate, because even in the changed invasion scenario most high-risk
boaters are expected to enter the province via the eastern border.

\section{Flexible and location-specific compliance rates\label{sec:Flexible-and-location-specific-compliance-rate}}

It may be more cost-effective to implement measures enforcing boaters'
compliance than to operate many inspection stations for long hours.
Furthermore, compliance of boaters may be higher or enforced more
easily at some specific locations. In this appendix, we show how the
approach presented in this paper can be adjusted to take these factors
into account.

\subsection{Location-specific compliance rates}

We start by considering the case of non-uniform compliance rates.
To that end, we split the boater flows based on the compliance of
the boaters. Let $C$ be the set of possible compliance rates, $c_{l}\in C$
the expected compliance rate of boaters at location $l\in L$, and
$L_{c}$ the set of locations with compliance rate $\tilde{c}\geq c$.
For a route $r\in R$ and a time interval $t\in T$ Let $n_{rtc}$
be the expected number of boaters who travel along route $r\in R$,
started their journey in time interval $t\in T$, and comply at all
inspection locations $l$ with $l_{c}\geq c$ but not at inspection
locations with $l_{c}<c$. These boaters will be inspected if and
only if 
\begin{eqnarray}
\smo{l\in L_{r}\cap L_{c}\,}\smo{s\in S_{lrt}}x_{ls} & \geq & 1.
\end{eqnarray}
As in the main text, $x_{ls}$ is a binary variable denoting whether
inspections are conducted at location $l\in L$ in shift $s\in S$.
Consequently, the total number of inspected boaters is given by

\begin{eqnarray}
F_{\mt{loc\text{-}compliance}}\ap{\vv x} & := & \smo{c\in C}\smo{r\in R}\smo{t\in T}\min\left\{ 1,\smo{l\in L_{r}\cap L_{c}}\smo{s\in S_{lrt}}x_{ls}\right\} n_{rtc}.\label{eq:taget-fun-operation-compliance}
\end{eqnarray}
This function can be optimized with the same method discussed in the
main text. With a similar approach, time-dependent compliance rates
could be incorporated, too.

\subsection{Flexible compliance rates}

In some applications, the compliance rate may be altered at a specific
cost. If these costs can be expressed as a convex function of the
achieved compliance rate, a flexible compliance rate can be incorporated
in our model easily. Below, we consider for simplicity the base case
with a uniform compliance rate at all locations. Allowing location-specific
flexible compliance rates can be done by combining the two approaches
introduced in this appendix.

Let $n_{rt}$ be the expected number of boaters travelling on route
$r\in R$ and who started their journey in time interval $t\in T$.
Note that other than in the main text, compliance of these boaters
is not supposed. Altering equation (5) from the main text to
\begin{eqnarray}
F_{\mt{flex\text{-}compliance}}\ap{\vv x} & := & c\smo{r\in R}\smo{t\in T}\min\left\{ 1,\smo{l\in L_{r}}\smo{s\in S_{lrt}}x_{ls}\right\} n_{rt}\label{eq:taget-fun-operation-1}
\end{eqnarray}
accounts for the flexible compliance rate $c$.

Let us assume assume that the costs for enforcing a specific compliance
rate $c$ at a location $l\in L$ and during shift $s\in S$ are given
by the linear function 
\begin{eqnarray}
\mt{cost}_{ls}\ap c & = & \alpha_{l}\left(c-c_{0}\right),
\end{eqnarray}
whereby $c_{0}$ is the base compliance rate if no actions are taken
to increase compliance. More complex cost functions can be modelled
with convex piece-wise linear functions or general convex functions.
Adding these costs to the overall cost function changes the cost constraint
to

\begin{eqnarray}
\smo{l\in L}\left(\smo{s\in S_{l}}\left(c_{ls}^{\mt{shift}}+\alpha_{l}\left(c-c_{0}\right)\right)x_{ls}+c_{l}^{\mt{loc}}\maxo{r\in R,\,t\in T}\left(\smo{s\in S_{lrt}}x_{ls}\right)\right) & \leq & B.\label{eq:cost-constr-operation-1}
\end{eqnarray}

In addition to changing the objective function and the cost constraint,
we have to introduce one further constraint limiting the compliance
rate to the feasible range:
\[
c_{0}\la{\leq}c\la{\leq}1.
\]
With these changes, the compliance rate can be optimized along with
the inspection locations and operating times.

\section{Difficult inspection optimization scenarios\label{sec:Difficult-inspection-optimizatio-Scenarios}}

In many real-world instances, good solutions to the linear integer
problems derived in this paper can be identified within reasonable
time. Nonetheless there are examples in which the optimization is
computationally challenging. In this appendix, we discuss two important
mechanism that can make it difficult to find a highly optimal solution
in short time. We also provide examples for the discussed mechanisms.

Difficulties can arise (1) if a significant fraction of the budget
is unused under the optimal policy and (2) if many boaters pass multiple
operated inspection locations under the optimal policy. We start by
considering budget-related issues before we discuss problems arising
from unfavourable relationships between potential inspection locations.
At the end of this appendix we discuss why these challenges are not
of major concern in many real-world applications. To simplify explanations,
we consider the case of optimizing inspection station placement only.
The described mechanisms extend easily to the full problem in which
operating times must be optimized as well.

\subsection{Difficulties due to cost constraints}

Let us first consider a scenario in which a fraction of the given
budget remains unused under the optimal policy. For example, suppose
that operation of an inspection station costs $5$ cost units and
that we are given a budget of $9$ units. Consequently, $4$ cost
units of the budget will remain unused. To obtain an approximate solution
and obtain an upper bound on the optimal objective value, solvers
consider the problem's continuous relaxation, in which partial use
of inspection locations (and shifts) is permissible. In this relaxed
scenario, all $9$ cost units will be spent, which allows the inspection
of more boaters than in the realistic scenario with binary choices.
Consequently, the upper bound on the solution given by the solution
to the relaxed problem may be much higher than the true optimal solution.
This makes it difficult to check whether an identified solution is
highly optimal and thus increases computation time.

The problem described above becomes even more difficult if control
actions with different costs are possible. Suppose that we may operate
one of three inspection stations, which are passed by different sets
of boaters, respectively. That is, no boater passes two of the potential
inspection locations. Assume that per day $n_{1}=n_{2}=5$ boaters
pass stations $l_{1}$ and $l_{2}$, respectively, and that $n_{3}=8$
boaters may be inspected at location $3$. Suppose we are given a
budget of $9$ units and that the costs for operating stations $l_{1}$
and $l_{2}$ are $c_{1}=c_{2}=5$ cost units, whereas operation of
station $l_{3}$ requires $c_{3}=9$ cost units.

Again, optimizers may consider the problem's continuous relaxation
to find an approximate solution and a quality estimate. An optimal
solution to the relaxed problem is to operate both station $1$ and
station $2$ fractionally with weight $x_{1}=x_{2}=0.9$. Then, the
total costs $x_{1}c_{1}+x_{2}c_{2}=9$ satisfy the budget constraint
and the total number of inspected boaters is given by $x_{1}n_{1}+x_{2}n_{2}=9$.
However, in the original integer problem, stations cannot be operated
fractionally, and only one station can be chosen. As more boaters
pass location $3$ than locations $l_{1}$ or $l_{2}$, it would be
optimal to conduct inspections at location $l_{3}$, where $8$ boaters
can be inspected. Applying a greedy rounding algorithm to the relaxed
solution, however, would suggest to operate either location $l_{1}$
or $l_{2}$, where only $5$ boaters would be expected.

\subsection{Difficulties due to unfavourable relations between inspection locations}

\begin{figure}
\begin{centering}
\includegraphics[scale=0.9]{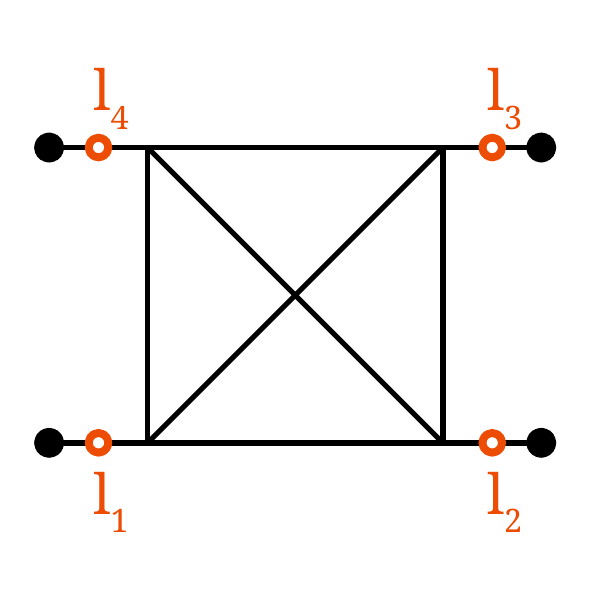}
\par\end{centering}
\caption[Inspection location setup that leads to a challenging optimization
problem]{Inspection location setup that leads to a challenging optimization
problem. The lines denote roads, the solid black circles origins and
destination, and the hollow orange circles potential inspection locations.\label{fig:Inspection-location-setup-examples}}
\end{figure}

Besides challenges induced by cost constraints, specific relationships
between potential inspection locations can make the optimization difficult.
Consider the example depicted in Figure \ref{fig:Inspection-location-setup-examples},
whereby an arbitrary number of boaters may drive from each origin/destination
(black circle) to each other origin/destination. Suppose that operating
an inspection location at any of the permissible locations has unit
cost and that we are provided a budget of $2$ cost units. If the
relaxed version of the problem is considered and fractional operation
of stations is permitted, operating each location with intensity $\frac{1}{2}$
would cover all boater flows and hence be the optimal solution. However,
if discrete choices must be made, some boaters will not be inspected.
As all locations are operated equally in the optimal solution to the
relaxed problem, this solution does not provide any hint towards which
of the locations should be operated in the original scenario with
binary decisions. Therefore, the problem is difficult to solve.

\subsection{Prevalence of difficult scenarios in real-world applications}

Any of the challenging scenarios discussed above can occur in real-world
problems. However, certain characteristics of real-world scenarios
lower the risk of running into optimization issues. In many management
scenarios of interest, various inspection stations can be operated.
Problems induced by the budget constraint become less significant
if a large budget is considered so that a potential remainder of the
budget becomes insignificant. For example, in all scenarios with a
budget above $30$ units considered in this paper, we reached a solution
with at least $98\%$ optimality within minutes. Furthermore, issues
induced by budget constraints can be mitigated by investigating alternative
scenarios with slightly adjusted budgets.

Scenarios with unfavourable relationships between potential inspection
locations can be expected in real-world applications. Note that the
issue with the setup in Figure \ref{fig:Inspection-location-setup-examples}
persists if the roads connecting the potential inspection locations
have a shape different from the road pattern drawn in the figure.
Furthermore, the depicted situation may refer to a portion of the
road network only, with multiple origins and destinations connected
to each of the depicted origin/destination vertices. In fact, situations
such as the considered one could appear multiple times in a road network.
Therefore, the considered challenges do not only occur in scenarios
in which inspections are restricted to locations close to origins
and destinations.

Nonetheless, invasion patterns frequent in real-world scenarios reduce
the prevalence of such unfavourable inspection station relationships.
As short distance dispersal of invasive species is typically more
likely than long-distance dispersal, invaded habitat patches form
clusters so that the inflow of potentially infested vectors, such
as boaters, comes from specific directions only. For example, high-risk
boaters enter BC through the southern and eastern border only. Therefore,
it is often possible to identify inspection location configurations
in which only few high-risk boaters pass multiple operated inspection
stations. This simplifies optimization of the inspection policy. Greater
optimization challenges can be expected if origins and destinations
are intermixed.